# Daily Forecasting of New Cases for Regional Epidemics of Coronavirus Disease 2019 with Bayesian Uncertainty Quantification

Yen Ting Lin[1,*], Jacob Neumann[2], Ely F. Miller[2], Richard G. Posner[2], Abhishek Mallela[3], Cosmin Safta[4], Jaideep Ray[4], Gautam Thakur[5], Supriya Chinthavali[5], and William S. Hlavacek[1]

[1]*Los Alamos National Laboratory, Los Alamos, New Mexico, USA*

[2]*Northern Arizona University, Flagstaff, Arizona, USA*

[3]*University of California, Davis, California, USA*

[4]*Sandia National Laboratories, Livermore, California, USA*

[5]*Oak Ridge National Laboratory, Oak Ridge, Tennessee, USA*

To increase situational awareness and support evidence-based policy-making, we formulated two types of mathematical models for COVID-19 transmission within a regional population. One is a fitting function that can be calibrated to reproduce an epidemic curve with two timescales (e.g., fast growth and slow decay). The other is a compartmental model that accounts for quarantine, self-isolation, social distancing, a non-exponentially distributed incubation period, asymptomatic individuals, and mild and severe forms of symptomatic disease. Using Bayesian inference, we have been calibrating our models daily for consistency with new reports of confirmed cases from the 15 most populous metropolitan statistical areas in the United States and quantifying uncertainty in parameter estimates and predictions of future case reports. This online learning approach allows for early identification of new trends despite considerable variability in case reporting. We infer new significant upward trends for five of the metropolitan areas starting between 19-April-2020 and 12-June-2020.

**Keywords:** Mathematical Model, Statistics, Uncertainty, Epidemics, Coronavirus Disease 2019 (COVID-19), Severe Acute Respiratory Syndrome Coronavirus 2 (SARS-CoV-2)

Coronavirus disease 2019 (COVID-19), which is caused by severe acute respiratory syndrome coronavirus 2 (SARS-CoV-2) *(1)*, was detected in the United States (US) in January 2020 *(2)*. In February, COVID-19-caused deaths were detected *(3)*. Thereafter, surveillance testing expanded nationwide *(4)*. These and other efforts revealed community spread across the US and exponential growth of new COVID-19 cases throughout most of March with a doubling time of 2 to 3 d *(5)*, similar to that of the initial outbreak in China *(6)*. This situation led to government mandates closing schools, prohibiting public gatherings, and restricting commercial activities, as well as broad adoption of social-distancing practices, such as working-from-home, curtailing of travel, and face mask-wearing *(7)*. Although the US became a hot spot of the COVID-19 pandemic, detection of new cases peaked in late-April and steadily declined until mid-June *(4)*, suggesting that public-health mandates and social-distancing practices were effective at slowing COVID-19 transmission. Attempts to quantify the impacts of these measures suggest substantial benefits in terms of reduction of disease burden *(8, 9)*.

At present, the number of new daily cases in the US is again increasing *(4)*. It is imperative that we effectively monitor ongoing COVID-19 transmission, so that dangerous upticks in cases can be responded to as quickly as possible.

To contribute to situational awareness of COVID-19 transmission dynamics, we developed two distinct mathematical models for the regional COVID-19 epidemic in each of the 15 most populous US metropolitan statistical areas (MSAs) *(10)*. The two models are 1) a fitting function, which can reproduce the shape of an epidemic curve having two timescales (e.g., fast growth and slow decay), and 2) a compartmental model. The latter model is composed of ordinary differential equations (ODEs) characterizing the population dynamics of susceptible, presymptomatic, symptomatic, asymptomatic, and recovered populations and various subpopulations, including mixing and protected-by-social-distancing populations, individuals who are quarantined or self-isolated, individuals incubating virus without symptoms at multiple stages of disease progression, and individuals with mild and severe forms of disease. The model also tracks hospitalizations and deaths.

In an ongoing online-learning effort, we are calibrating our models for regional COVID-19 epidemics on a daily basis for consistency with historical case reports. We have also been applying Bayesian methods to quantify uncertainties in predicted detection of new cases. In the face of variability in case detection, this approach allows for early objective identification of new epidemic trends beyond what can be reasonably explained by a well-calibrated model that is consistent with historical case reporting, enabling an evidence-based response by policy makers.

**Methods**

*Data Used in Online Learning*

The COVID-19 surveillance data used to parameterize models—reports of new confirmed cases, the earliest widely available indicators of local trends—are obtained daily (at variable times of day) from the GitHub repository maintained by *The New York Times* newspaper *(11)*. This repository aggregates reports from US State and local health agencies. We aggregate county-level case counts to obtain case counts for each of the 15 most populous US MSAs. These MSAs encompass the following cities: New York City; Los Angeles; Chicago; Dallas; Houston; Washington, DC; Miami; Philadelphia; Atlanta; Phoenix; Boston; San Francisco; Riverside, CA; Detroit; and Seattle. We will use these city names to refer to the MSAs. The political entities comprising each MSA, which are almost always counties[1], are those delineated by the federal government *(10)*.

*Models for COVID-19 Transmission and Model Parameters*

In ongoing work, each day, for the regional COVID-19 epidemic in each of the 15 MSAs of interest, we parameterize a curve-fitting model—a fitting function—and a compartmental model for consistency with all daily reports of new confirmed cases that are available at the time.

The form of the curve-fitting model is defined by Equations (2)–(4) in the Appendix. This form was chosen because of its ability to generate asymmetrically shaped curves (Figure 1). The curve-fitting model is taken to have four adjustable parameters (Table 1): $N$, the total number of infected individuals who will be detected over the entire course of the local epidemic; $t_0$, the start time of the local epidemic; and $k$ and $\theta$, the shape and scale parameters of a gamma ($\Gamma$) distribution. Inference of adjustable parameter values is based on a negative binomial likelihood function, given by Equation (31) in the Appendix. The dispersal parameter $r$ of the likelihood is taken to be adjustable; its value is jointly inferred with those of $N$, $t_0$, $k$, and $\theta$. The inferred parameter values are inference-time-dependent and region-specific. Inferences are conditioned on the model and fixed parameter estimates for $\mu$ and $\sigma$ (Table 1), which characterize the waiting time from infection to onset of symptoms *(12)*, which is taken to be a log-normally distributed random variable.

The compartmental model was formulated to capture populations and processes important in COVID-19 transmission (Figure 2). Appendix Figure 1 provides a more detailed illustration of the model. In the model, the susceptible (S), exposed (E), infectious (I), and removed (R) populations of the classic SEIR model are all considered but are divided into subpopulations. An important feature of the model, introduced to capture the effects of social distancing, is that susceptible and infectious individuals are divided into mixing and protected subpopulations. In the mixing population, individuals interact with others as they would normally (i.e., without taking special precautions to prevent COVID-19 transmission), whereas in the protected population, individuals practice social distancing, which is taken to slow COVID-19 transmission. The model consists of 25 ordinary differential equations (ODEs), defined by Equations (5)–(26) in the Appendix. Each state variable of the model represents the size of a population (or in other terminology, the population of a compartment). In addition to the 25 ODEs, we consider an auxiliary 1-parameter measurement model, which relates state variables to expected case reporting (see Equations (27) and (28) in the Appendix). The model is formulated so as to allow consideration of multiple periods of social distancing with distinct setpoints for the protected population size. In the model, there is always an initial period of social distancing. The number of additional social-distancing periods is indicated by $n$. Here, we only consider two cases: $n = 0$ and $n = 1$.

For $n = 0$ (i.e., only one social-distancing period), the compartmental model and auxiliary measurement model have a total of 20 parameters. We take six of these parameters to have adjustable values (Table 2) and 14 to have fixed values (Table 3). In the Appendix, we describe each parameter and explain the fixed parameter settings, which are based on information in Refs. *(12)–(20)* and assumptions. The adjustable model parameters are as follows: $t_0$, the start time of the local epidemic; $\sigma > t_0$, the time at which social distancing begins; $p_0$, which establishes a setpoint for the quasi-stationary fraction of the total population practicing social distancing; $\lambda_0$, which establishes a timescale for adoption of social-distancing practices; and $\beta$, which characterizes the rate of disease transmission in the absence of social distancing. The measurement-model parameter, $f_D$, represents the time-averaged fraction of new cases detected, which is related to the intensity of local surveillance efforts. As for the curve-fitting model, inference of adjustable parameter values is based on a negative binomial likelihood function (Equation (31) in the Appendix). The dispersal parameter $r$ of the likelihood is taken to be adjustable; its value is jointly inferred with those of $t_0$, $\sigma$, $p_0$, $\lambda_0$, $\beta$, and $f_D$. All inferred parameter values are inference-time-dependent and region-specific. Inferences are conditioned on the model and fixed parameter estimates.

For $n > 0$ (i.e., $n$ distinct social-distancing periods following an initial or first-phase social-distancing period), the compartmental model has three additional adjustable parameters for each additional period of social distancing considered beyond the initial period. For one additional period of social distancing ($n = 1$), the additional adjustable parameters are as follows: $\tau_1 > \sigma$, the onset time of second-phase social-distancing; $p_1$, the second-phase quasi-stationary setpoint parameter; and $\lambda_1$, which determines the timescale for transition from first- to second-phase social-distancing behavior. If adherence to effective social-distancing practices begins to relax at time $t = \tau_1$, then $p_1 < p_0$.

*Statistical Model for Noisy Case Reporting*

The curve-fitting and compartmental models are both deterministic, despite COVID-19 transmission being a stochastic process. We interpret each model to predict the expected number of new confirmed COVID-19 cases reported daily by the relevant public health authorities. In other words, we assume that the number of new cases reported over a 1-d period is a random variable and its expected value follows a deterministic trajectory. We further assume that day-to-day fluctuations in the random variable are independent and characterized by a negative binomial distribution, which we will denote as $\text{NB}(r, p)$. We use $\text{NB}(r, p)$ to statistically model noise in reporting (and case detection, which influences reporting) because its support—the non-negative integers—is natural for populations and, in addition, its shape is flexible enough to recapitulate a wide range of unimodal empirical distributions. With these assumptions, we obtain a likelihood function (Equation (31) in the Appendix) taking the form of a product of probability mass functions of $\text{NB}(r,p)$. Formulation of a likelihood is a prerequisite for standard Bayesian inference[2].

*Online Learning of Model Parameter Values through Bayesian Inference*

We use Bayesian inference to learn adjustable model parameter values consistent with the time-series of daily MSA-specific reports of new COVID-19 cases available up to the point of inference. Inferences with both models are performed daily for each MSA of interest. In each inference, we assume a uniform prior and use an adaptive Markov chain Monte Carlo (MCMC) algorithm *(21)* to generate samples of the posterior distribution for the adjustable parameters. A full description of our inference procedure is provided in the Appendix.

The maximum a posteriori (MAP) estimate of a parameter value is the value of the parameter corresponding to the mode of its marginal posterior, where probability mass is highest.

*Forecasting with Quantification of Prediction Uncertainty: Bayesian Predictive Inference*

In addition to inferring parameter values, we quantify uncertainty in predicted trajectories of daily case reports. A predictive inference of the expected number of new cases detected on a given day is derived from a model by parameterizing it using a randomly chosen parameter posterior sample generated in MCMC sampling. The number of these cases that are detected is then predicted by adding a noise term, drawn from a negative binomial distribution $\text{NB}(r,p)$, where $r$ is set at the randomly sampled value—recall that the value of $r$ is jointly inferred with adjustable model parameter values—and $p$ is set such that the mean of $\text{NB}(r,p)$ corresponds to the predicted number of new active cases, i.e., the value of $p$ is set using Equation (30) in the Appendix. A full description of our predictive inference procedure is provided in the Appendix.

Predictions of the curve-fitting model are obtained by evaluating the sum in Equation (2) in the Appendix. Predictions of the compartmental model are obtained by using LSODA *(22)* to numerically integrate Equations (5)–(21) and (27) in the Appendix; the initial condition is defined by the inferred value of $t_0$ (Table 2) and the fixed settings for $S_0$ and $I_0$ (Table 3).

The 95% credible interval for the predicted number of new case reports on a given day is the central part of the marginal predictive posterior capturing 95% of the probability mass. This region is bounded above by the 97.5 percentile and below by the 2.5 percentile

**Results**

The objective of our ongoing study is to detect significant new trends in new COVID-19 cases as early as possible by 1) systematically and regularly updating mathematical models capturing historical trends in regional COVID-19 epidemics through Bayesian inference and 2) making forecasts with rigorously quantified uncertainties through Bayesian predictive inference.

An important aspect of how we are analyzing COVID-19 data is our focus on the populations of US cities and their surrounding areas (i.e., MSAs) vs. the regional populations within other political boundaries, such as those of the US States. The boundaries of MSAs are defined on the basis of social and economic interactions *(10)*, which suggests that the population of an MSA is likely to be more uniformly affected by the COVID-19 pandemic than, for example, the population of a State. In accordance with this expectation, daily reports of new COVID-19 cases for New York City (Figure 3, panel A) are more temporally correlated than for the US State of New York (Figure 3, panel B), New Jersey (Figure 3, panel C), or Pennsylvania (Figure 3, panel D). New York, New Jersey, and Pennsylvania are the three States encompassing the New York City MSA.

For each of the 15 most populous MSAs in the US, we parameterized a curve-fitting model and a compartmental model using MSA-specific surveillance data, namely, aggregated County-level reports indicating the number of new confirmed COVID-19 cases within a given MSA each day. Bayesian parameterization and forecasting with uncertainty quantification (UQ)—predictive inference—are now being performed daily for each of the 15 MSAs, with both models.

Results of Bayesian predictive inference are exemplified in Figure 4. Predictions are obtained in the form of a predictive posterior distribution and are based on all case reports available at the time of inference. In Figure 4, the entire shaded region indicates the 95% credible interval of the predictive posterior as a function of time. In other words, this band indicates where 95% of predictions of daily case reports fall at the times indicated. Predictions vary because of the uncertainties in adjustable model parameter estimates, which are characterized quantitatively through Bayesian inference. The colors within the shaded band indicate other credible intervals (10%, 20%, etc.) and also the median of all predictions as a function of time.

Predictive inferences for all 15 MSAs of interest are shown in Figures 5 and 6. The predictions of Figure 5 are conditioned on the curve-fitting model, and the predictions of Figure 6 are conditioned on the compartmental model. These results demonstrate that the curve-fitting and compartmental models are each capable of reproducing empirical epidemic curves for multiple MSAs, which vary in shape. However, the curve-fitting model is less flexible, as can be seen by comparing its predictions to those of the compartmental model for the Atlanta MSA, where there is high variability in the daily number of new cases detected. Although there is no clear downward trend in the data, the curve-fitting model nevertheless predicts a peak in late-April/early-May and a downward trend ever since. This prediction is obtained because the model, by design, is only capable of generating single-peaked epidemic curves that rise and then fall. Hereafter, we will focus on the compartmental model.

Recall that predictive inferences are made daily. In Figure 7, we show predictive inferences for New York City and Phoenix made over a series of progressively later dates. These results illustrate that accurate short-term predictions are possible but continual updating of parameter estimates is required to maintain accuracy. In Videos 1 and 2, daily predictions for New York City and Phoenix, respectively, are shown as animations.

In practice, we find that the adjustable parameters of the compartmental model have identifiable values[3], meaning that their marginal posteriors are unimodal. This finding is illustrated in Figure 8, which displays a 7 × 7 matrix of 1- and 2-dimensional projections of the 7-dimensional MCMC posterior samples underlying predictive inferences for New York City. Plots of marginal posteriors are shown on the diagonal extending from top left to bottom right. The other 2-dimensional plots reveal correlations between pairs of parameter estimates (if any). The significance of identifiability is that, despite uncertainties in parameter estimates, we can expect predictive inferences of daily new-case reports to cluster around a central trajectory.

How does learning the region-specific adjustable parameter values of the compartmental model and a subsequent predictive inference (i.e., a forecast with UQ) improve situational awareness? In the vast majority of cases, when we forecast with UQ, the empirical new-case count for the day immediately following our inference (+1), and very often for each of several additional days, falls within the 95% credible interval of the predictive posterior. When the reported number of new cases falls outside the 95% credible interval and above the 97.5% percentile, we interpret this event, which we will call an *upward-trending rare event*, to have a probability of 0.0275 or less assuming the model is both explanatory (i.e., satisfactorily consistent with historical data) and predictive of the near future. If the model is predictive of the near future, the probability of two consecutive rare events is far smaller, less than 0.001. Thus, consecutive upward-trending rare events, which we will refer to as an *upward-trending anomaly*, can be reasonably taken as a sign that the model is not in fact predictive. Indeed, an anomaly suggests that the rate of COVID-19 transmission has increased beyond what can be explained by the model.

For New York City, anomalies are not seen, as illustrated in Figure 9, panel A. However, for Phoenix, there are several recent anomalies, which preceded rapid and sustained growth in the number of new cases reported per day in June (Figure 9, panel B).

In an attempt to explain the anomalies, we allowed the compartmental model to account for a distinct second social-distancing period, i.e., we increased the setting for $n$ from 0 to 1. With this change, the number of adjustable parameters increases from 7 to 10, because in the model, each distinct social-distancing period is characterized by 3 parameters, which must be inferred. Recall that these parameters are an onset time, rate parameter, and setpoint parameter. The value of the setpoint parameter determines the quasi-stationary fraction of the total regional population that is adhering to effective social-distancing practices. In the model, $p_0$ is the setpoint parameter of the initial social-distancing period starting at time $t = \sigma > t_0$, and $p_1$ is the setpoint parameter of a distinct subsequent social-distancing period starting at time $t = \tau_1 > \sigma$. As can be seen by comparing the plots in Figure 10, panels A and B, the compartmental model with two social-distancing periods (Figure 10, panel B) better explains the Phoenix surveillance data available than the compartmental model with just one social-distancing period (Figure 10, panel A)[4]. Furthermore, the MAP estimate for $p_1$ (~0.38) is less than that for $p_0$ (~0.49) (cf. panels C and D, Figure 10), and the marginal posteriors for these parameters are largely non-overlapping (Figure 10, panel D). These findings suggest that the recent increase in COVID-19 cases in Phoenix can be explained by relaxation in social distancing, which is quantified by our estimates for $p_0$ and $p_1$. The MAP estimate of the start time of the second period of social distancing corresponds to 24-May-2020[5]. Intriguingly, 8 of the 9 anomalies noted in Figure 10, panel B occurred after this period, with the first of these occurring on 02-June-2020.

We considered the hypothesis that a one-time event (viz., a mass gathering) generating 1000's of new infections might trigger a new upward trend in COVID-19 transmission. Simulations for New York City and Phoenix do not support this hypothesis (Appendix Figure 2).

Besides Phoenix, four other MSAs have recent trends explainable by relaxation of social distancing (Appendix Figure 3 and Appendix Table 1). Upward-trending anomalies were detected for these MSAs (Appendix Figure 4, panels A–D), but not for three of four other MSAs having epidemic curves consistent with sustained social distancing (Appendix Figure 4, panels E–H). Daily predictions for the MSAs considered in Appendix Figure 4 are shown in Videos 3–10.

**Discussion**

Daily online learning of model parameter values from real-time surveillance data is feasible for mathematical models for COVID-19 transmission. Furthermore, predictive inference of the daily number of new cases reported is feasible for the regional COVID-19 epidemics occurring in multiple US metropolitan areas. Our model-based analyses are ongoing and daily forecasts are being disseminated *(23)*.

We have suggested how our predictive inferences can be used to identify harbingers of future growth in COVID-19 transmission rate. Two consecutive upward-trending rare events (i.e., instances where the number of new cases reported is above the upper limit of the 95% credible interval of the predictive posterior) seem to indicate potential for increased transmission in the future. This signal of future growth is perhaps especially strong when anomalies are accompanied by increasing prediction uncertainty, as is the case for Phoenix (Figure 9, panel B).

We find that the recent increase in rate of transmission of COVID-19 in the Phoenix metropolitan area can be explained by a reduction in the percentage of the population adhering to effective social-distancing practices[6] (Figure 10, panel D), from ~49% to ~38%. Relaxation is inferred to have begun around 24-May-2020 (Figure 10, panel B). Recent upward trends in the rate of COVID-19 transmission in the Houston, Miami, San Francisco, and Seattle metropolitan areas can also be explained by relaxation of social distancing (Appendix Figure 3 and Appendix Table 1). These findings are qualitatively consistent with earlier studies indicating that social distancing is effective at slowing the transmission of COVID-19 *(7, 8)*, and encouragingly, they suggest that the future course of the pandemic is controllable, especially with accurate recognition of when stronger nonpharmaceutical interventions are needed to slow COVID-19 transmission.

**Acknowledgments**

Y.T.L. was supported by the Laboratory Directed Research and Development Program at Los Alamos National Laboratory (Project XX01); this support enabled early feasibility studies. Y.T.L., C.S., J.R., G.T., S.C., and W.S.H. were supported by the US Department of Energy (DOE) Office of Science through the National Virtual Biotechnology Laboratory, a consortium of national laboratories (Argonne, Los Alamos, Oak Ridge, and Sandia) focused on responding to COVID-19, with funding provided by the Coronavirus CARES Act. J.N., E.M., and R.G.P. were supported by a grant from the National Institute of General Medical Sciences of the National Institutes of Health (R01GM111510). A.M. was supported by the 2020 Mathematical Sciences Graduate Internship program, which is sponsored by the Division of Mathematical Sciences of the National Science Foundation. Computational resources used in this study included the following: the Darwin cluster at Los Alamos National Laboratory (LANL), which is supported by the Computational Systems and Software Environment subprogram of the Advanced Simulation and Computing program at LANL, which is funded by National Nuclear Security Administration of the DOE and Northern Arizona University's Monsoon computer cluster, which is funded by Arizona's Technology and Research Initiative Fund.

**Disclaimers**

None.

**Author Bio**

Dr. Lin works as a Scientist in the Information Sciences Group of the Computer, Computational, and Statistical Sciences Division at Los Alamos National Laboratory. His primary research interest is development and application of advanced data-science methods in modeling of biological systems.

**Footnotes**

[1] For the metropolitan statistical areas (MSAs) of interest, the number of political units (viz., counties and independent cities) comprising an MSA ranges from 2 (for the Los Angeles and Riverside MSAs) to 29 (for the Atlanta MSA); the median (mean) number of Counties is 7 (10). The number of States encompassing an MSA ranges from 1 (for eight of the 15 MSAs) to 4 (for Philadelphia); the median (mean) number of encompassing States is 1 (2).

[2] It should be noted that some related methods, typified by approximate Bayesian computation (ABC), do not rely on a likelihood.

[3] We do not have a mathematical proof of identifiability.

[4] This conclusion is supported by the Akaike information criterion (AIC) and Bayesian information criterion (BIC) values we calculated for the two scenarios (Appendix Table 1). Although AIC and BIC are crude model-selection tools, because the posteriors here are non-normal, we deem them to be adequately discriminatory. Each strongly indicates that the model with two social-distancing periods is more explanatory of the data than the model with just one social-distancing period.

[5] The 95% credible interval places the start date within the period beginning on 20-May-2020 and ending on 28-May-2020.

[6] Unfortunately, our study sheds no light on which social-distancing practices are effective at slowing COVID-19 transmission.

Table 1. Parameters of the curve-fitting model ($N$, $t_0$, $k$, $\theta$, $\mu$, and $\sigma$) and the associated likelihood function ($p$ and $r$) used in predictive inference.

| Parameter | Estimate (Units) | Comment |
|---|---|---|
| $N$ | 470,000* | Population size |
| $t_0$ | 35* (d) | Start of COVID-19 transmission |
| $k$ | 6.6* | Shape parameter of $\Gamma(k, \theta)$ |
| $\theta$ | 7.9* | Scale parameter of $\Gamma(k, \theta)$ |
| $\mu$ | 1.6** | $\mu$-parameter of log-normal distribution |
| $\sigma$ | 0.42** | $\sigma$-parameter of log-normal distribution |
| $p$ | Constrained*** | Probability parameter of $\mathrm{NB}(r, p)$ |
| $r$ | 4.4* | Dispersal parameter of $\mathrm{NB}(r, p)$ |

*Estimates of the adjustable parameters ($N$, $t_0$, $k$, $\theta$, and $r$) are region-specific and inference-time-dependent. Inferences are performed daily. Here, we report the maximum a posteriori (MAP) estimates inferred for New York City using all confirmed COVID-19 case-count data available in the GitHub repository maintained by *The New York Times* newspaper *(11)* for the period starting on 21-January-2020 and ending on 21-June-2020 (inclusive dates). Time $t = 0$ corresponds to 0000 hours on 21-January-2020. **Estimates of the fixed parameters $\mu$ and $\sigma$ are those of Lauer et al. *(12)*. These parameter estimates define a log-normal distribution that reproduces the empirical distribution of waiting times for the onset of symptoms after infection with SARS-CoV-2. ***The value of $p$ is constrained, i.e., its reporting-time-dependent value is determined by a formula, which is given by Equation (30) in the Appendix.

Table 2. Inferred values of the adjustable parameters of the compartmental model ($t_0$, $\sigma$, $p_0$, $\lambda_0$, and $\beta$), the auxiliary measurement model ($f_D$), and the associated statistical model for noise in case detection and reporting ($r$).

| Parameter | Estimate* (Units) | Comment |
|---|---|---|
| $t_0$ | 33 (d) | Start of COVID-19 transmission |
| $\sigma$ | 33 (d) | Start of social distancing |
| $p_0$ | 0.87 | Social-distancing setpoint |
| $\lambda_0$ | 0.10 (/d) | Social-distancing rate parameter |
| $\beta$ | 2.0 (/d) | Disease-transmission rate parameter |
| $f_D$ | 0.12 | Fraction of active cases reported |
| $r$ | 12 | Dispersal parameter of $NB(p,r)$** |

*All estimates are region-specific and inference-time-dependent. Inferences are performed daily. Here, we report the maximum a posteriori (MAP) estimates inferred for the New York City MSA using all confirmed COVID-19 case-count data available in the GitHub repository maintained by *The New York Times* newspaper *(11)* for the period starting on 21-January-2020 and ending on 21-June-2020 (inclusive dates). Time $t = 0$ corresponds to 0000 hours on 21-January-2020. **The probability parameter of $NB(r,p)$ is constrained, i.e., its reporting-time-dependent value is determined by a formula, which is given by Equation (30) in the Appendix.

Table 3. Estimates for the fixed parameters of the compartmental model.

| Parameter | Estimate (Units) | Source |
|---|---|---|
| $S_0$ | 19,216,182* | (13) |
| $I_0$ | 1 | Assumption |
| $n$ | 0** | Assumption |
| $m_b$ | 0.1 | Assumption |
| $\rho_E$ | 1.1 | Arons et al. (14) |
| $\rho_A$ | 0.9 | Nguyen et al. (15) |
| $k_L$ | 0.94 (/d) | Lauer et al. (12) |
| $k_Q$ | 0.0038 (/d) | Assumption |
| $j_Q$ | 0.4 (/d) | Assumption |
| $f_A$ | 0.44 | (16, 17) |
| $f_H$ | 0.054 | Perez-Saez et al. (18) |
| $f_R$ | 0.79 | Richardson et al. (19) |
| $c_A$ | 0.26 (/d) | Sakurai et al. (17) |
| $c_I$ | 0.12 (/d) | Wölfel et al. (20) |
| $c_H$ | 0.17 (/d) | Richardson et al. (19) |

*All estimates listed in this table are taken to apply to all regions of interest except for $n$, the number of distinct social-distancing periods that follow an initial social-distancing period, and $S_0$, the region-specific initial number of susceptible individuals. The value given here for $S_0$ is the US Census Bureau-estimated total population of the New York City metropolitan statistical area. **We assume $n = 0$ unless stated otherwise.

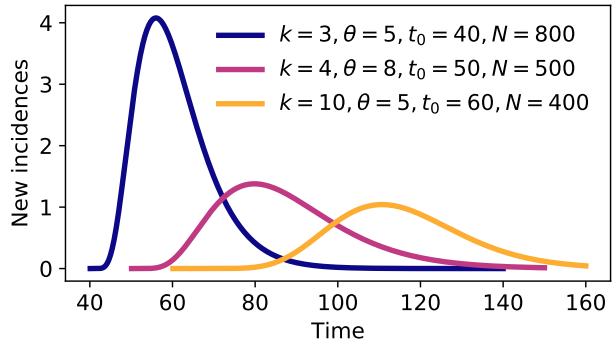

Figure 1. Illustration of shapes that can be produced by the fitting function that we are using to capture trends in regional COVID-19 epidemic curves. The curve-fitting model is formulated such that it has the capacity to reproduce the shape of an epidemic curve having two timescales.

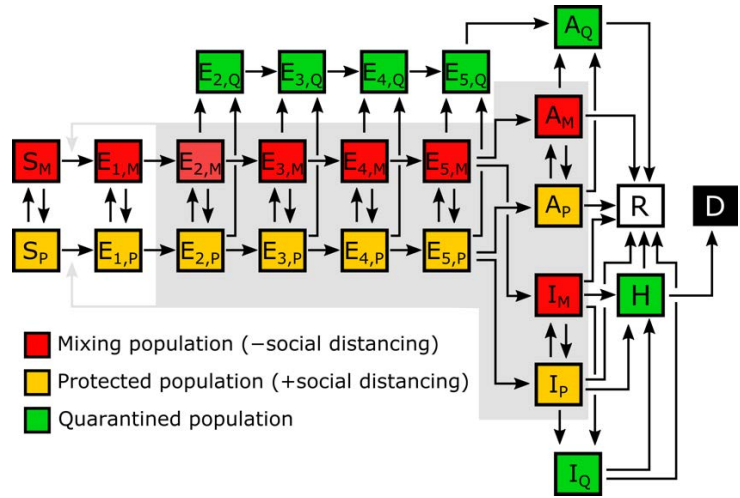

Figure 2. Illustration of the populations and processes considered in a mechanistic compartmental model for the dynamics of COVID-19 transmission. The model accounts for susceptible (S), exposed (E), asymptomatic (A), symptomatic (I), hospitalized (H), recovered (R), and deceased (D) populations. It also accounts for social distancing, which establishes mixing and protected subpopulations, quarantine driven by testing and contact tracing, and self-isolation spurred by symptom awareness. The incubation period is divided into 5 stages, which allows the model to reproduce an empirically determined Erlang distribution of waiting times for the onset of symptoms after infection *(12)*. The exposed population (consisting of individuals incubating virus) includes presymptomatic and asymptomatic individuals. The *A*-populations consist of true asymptomatic individuals in the immune clearance phase. The gray background indicates the populations that contribute to disease transmission. An auxiliary measurement model (Equations (27) and (28) in the Appendix) accounts for imperfect detection and reporting of new cases.

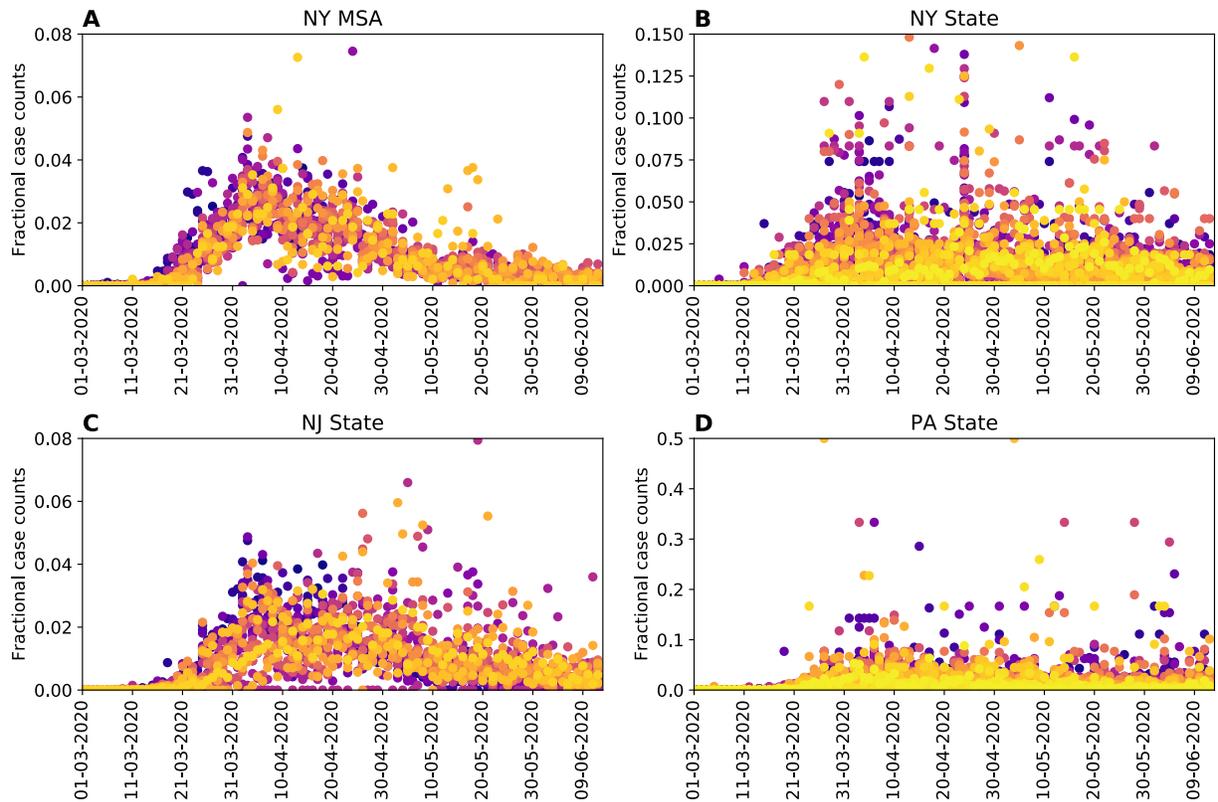

Figure 3. Temporal correlations in surveillance data. Shown here are time-series of fractional (i.e., normalized) case counts. We define the fractional case count for a county on a given date to be the reported number of cases on that date divided by the total reported number of cases in the county over the entire period of interest. The panels in this figure show fractional case counts for **(A)** the 23 counties comprising the New York City metropolitan statistical area (MSA), **(B)** the 62 counties comprising the State of New York, **(C)** the 21 counties comprising the State of New Jersey, and **(D)** the 67 counties comprising the State of Pennsylvania. Within each plot, a different color is used for the data points from each distinct county. As can be seen, time-series for the counties of the New York City MSA are more temporally correlated than for the State-level time-series. Daily case counts for New Jersey are similar to those for New York City because the two populations overlap considerably: ~74% of New Jersey's population is part of the New York City MSA and ~32% of the population of the New York City MSA is part of the State of New Jersey.

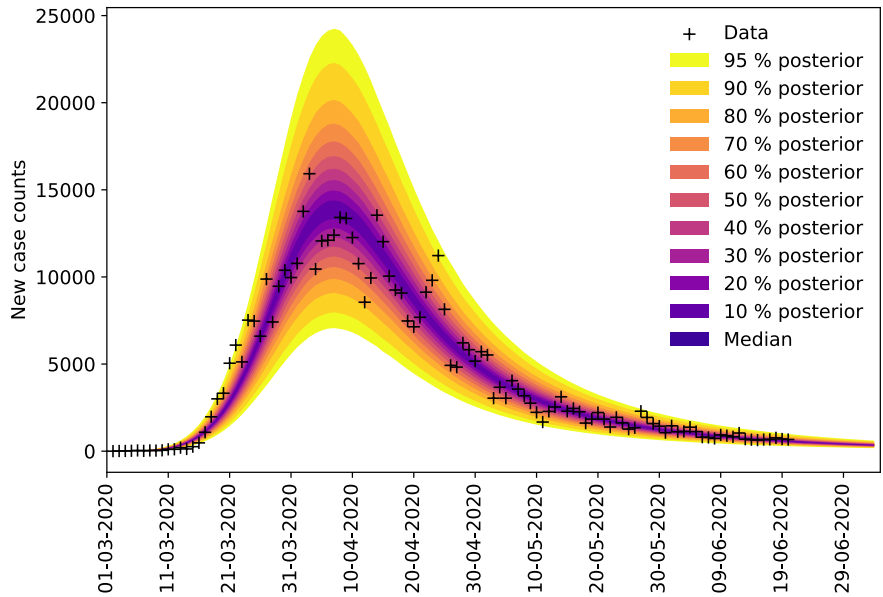

Figure 4. Illustration of Bayesian predictive inference. We forecast future daily reports of new COVID-19 cases with rigorous uncertainty quantification (UQ) through online Bayesian learning of model parameters. Each day, using all daily case-reporting data available up to that point, we perform Markov chain Monte Carlo (MCMC) sampling of the posterior distribution for a set of adjustable parameters. Subsampling of the posterior samples then allows us to use the relevant model to generate trajectories of the epidemic curve that account for both parametric and observation uncertainty. The entire shaded region indicates the 95% credible interval for predictions of daily case reports. In other words, the central 95% of all predictions lie within the shaded region. The color-coded bands within the shaded region indicate other credible intervals, as indicated in the legend.

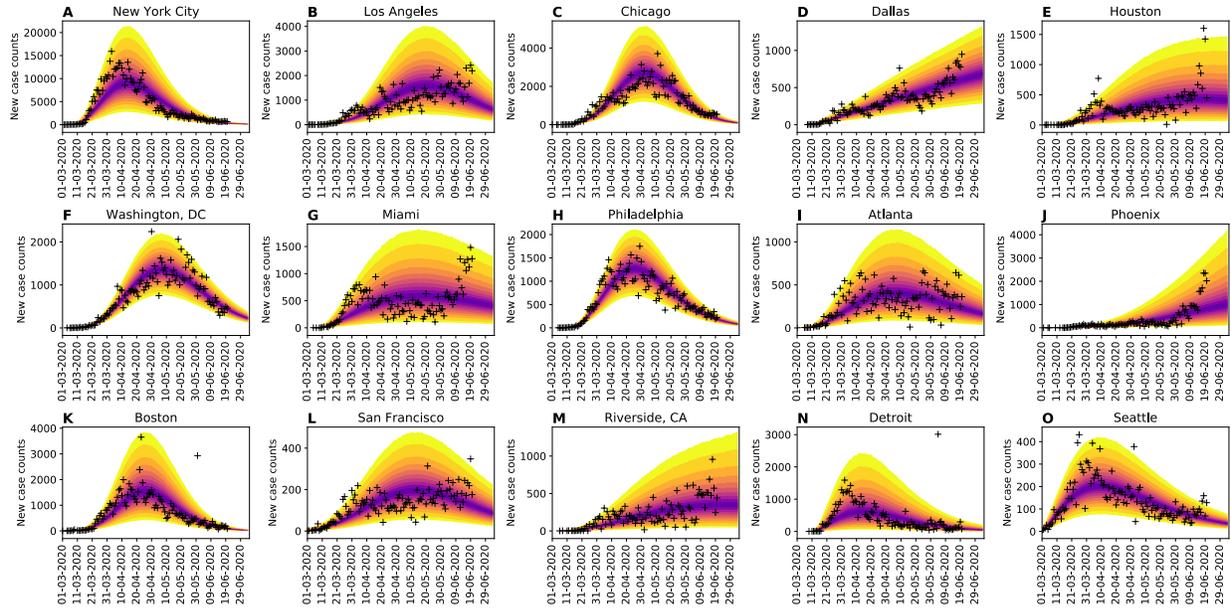

Figure 5. Bayesian predictive inferences for the 15 most populous metropolitan statistical areas (MSAs) in the United States. Predictions are conditioned on the curve-fitting model.

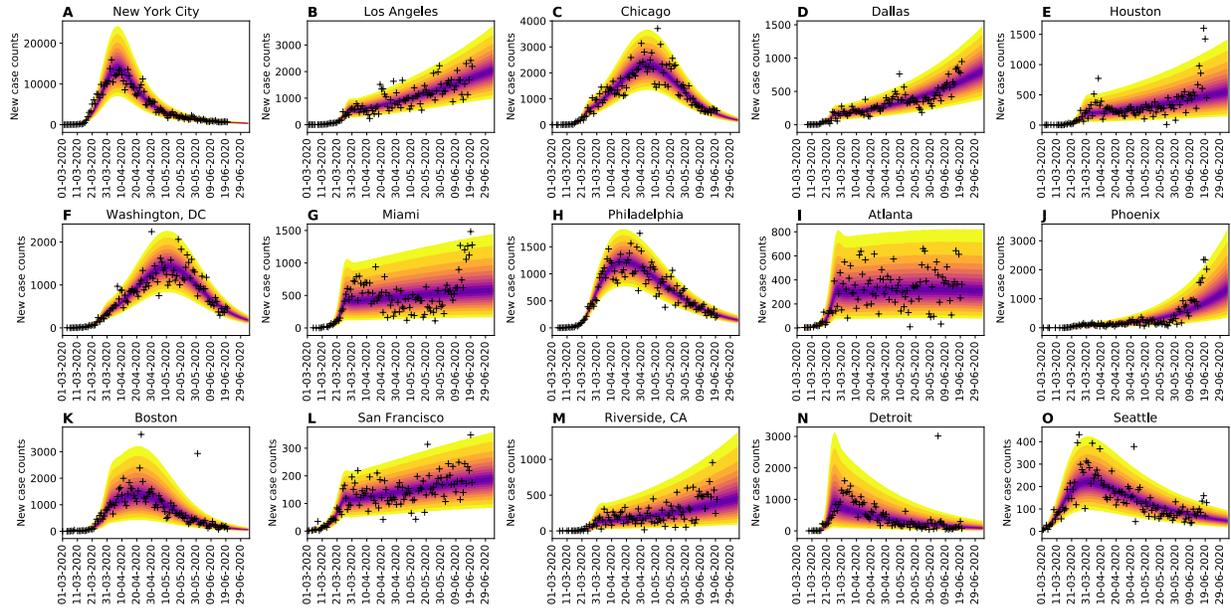

Figure 6. Bayesian predictive inferences for the 15 most populous metropolitan statistical areas (MSAs) in the United States. Predictions are conditioned on the compartmental model.

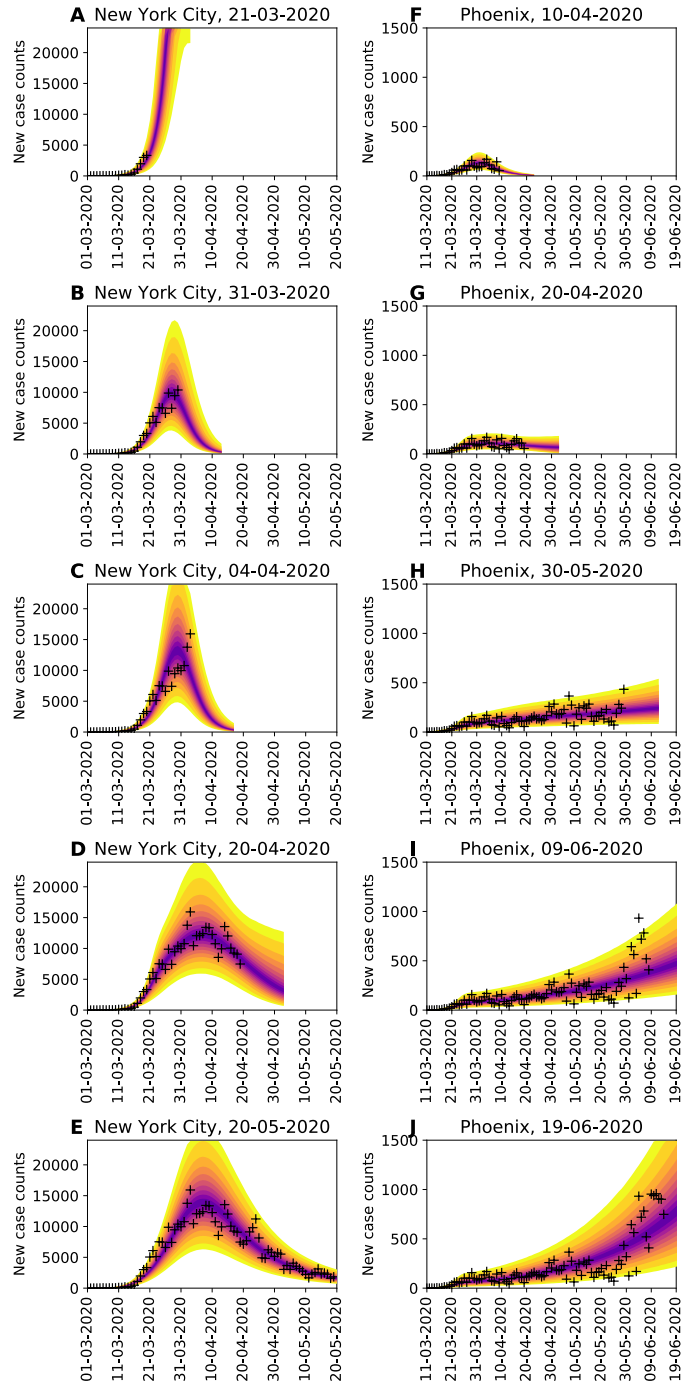

Figure 7. The necessity of online learning. (A)–(E) Shown are predictions for the New York City metropolitan statistical area (MSA) made over a series of progressively later dates, as indicated. (F)–(J) Shown are predictions for the Phoenix MSA made over a series of progressively later dates, as indicated. Predictive inferences, which are all conditioned on the compartmental model, are data-driven. Accurate short-term predictions are possible but continual updating of parameter estimates is required to maintain accuracy.

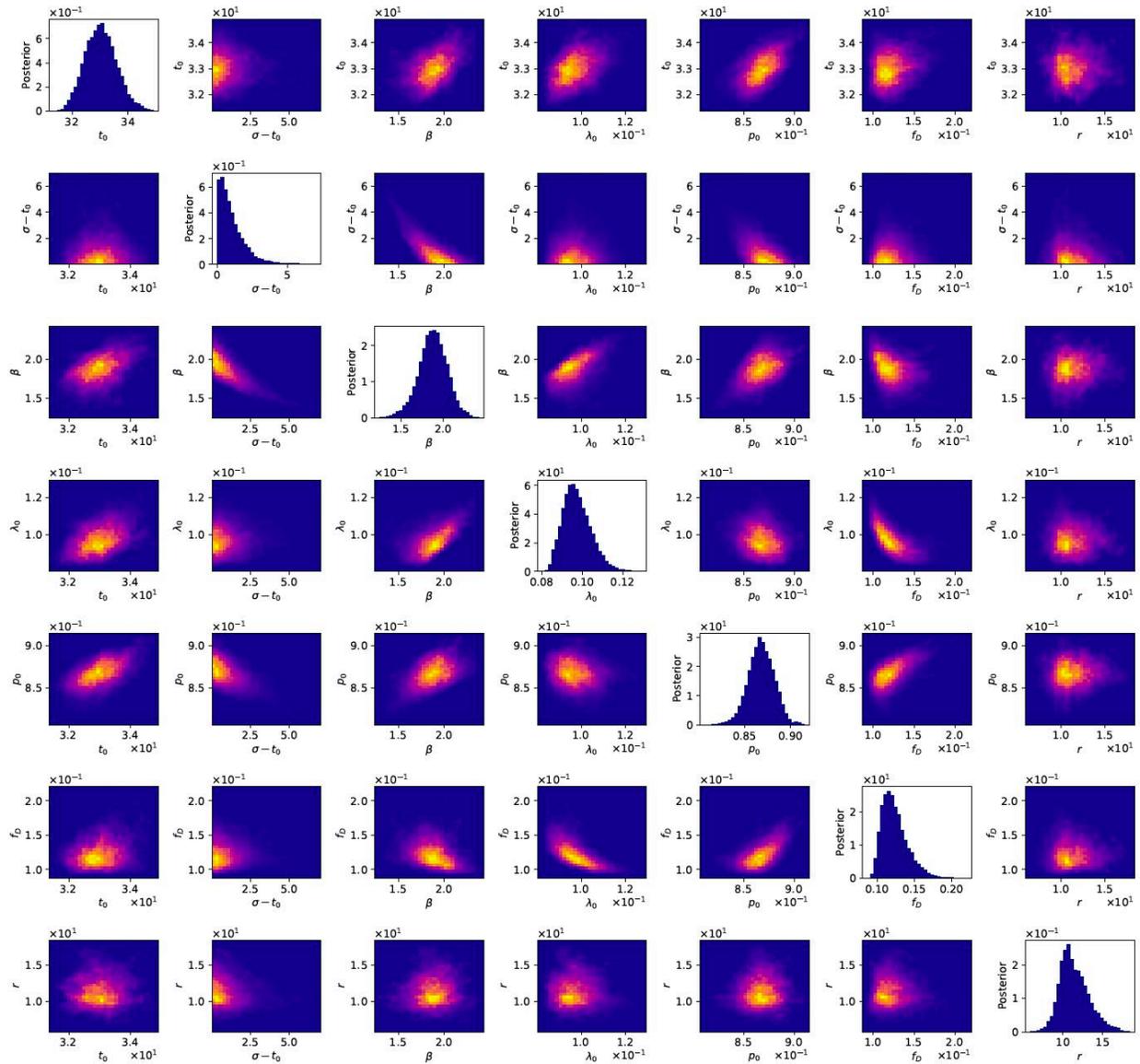

Figure 8. Matrix of 1- and 2-dimensional projections of the 7-dimensional posterior samples obtained for the adjustable parameters associated with the compartmental model for the New York City metropolitan statistical area (MSA) on the basis of daily reports of new confirmed coronavirus disease 2019 (COVID-19) cases from 21-January-2020 to 21-June-2020 (inclusive dates). Plots of marginal posteriors (1-dimensional projections) are shown on the diagonal from top left to bottom right. Other plots are 2-dimensional projections, which indicate how correlated pairs of parameter estimates are. Brightness indicates higher probability density. A compact bright area indicates absence of or relatively low correlation. An extended, asymmetrical bright area indicates relatively high correlation.

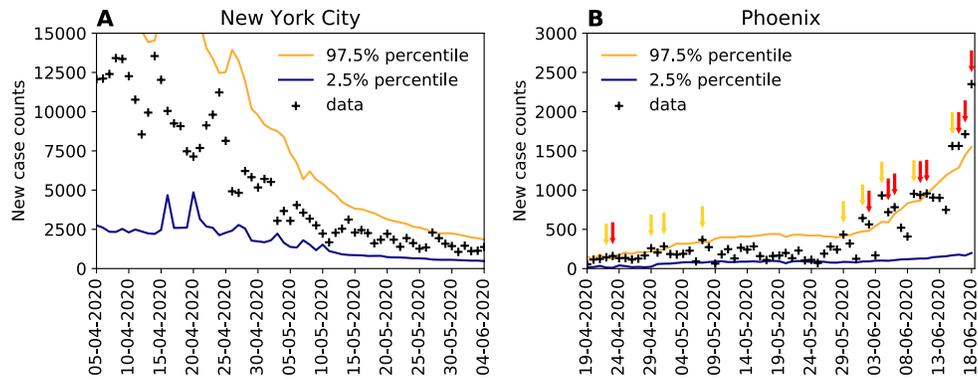

Figure 9. Rare events and anomalies, as defined in the main text, detected in the surveillance data available for **(A)** the New York City metropolitan statistical area (MSA) and **(B)** the Phoenix MSA. Yellow arrows mark upward-trending rare events. Red arrows mark upward-trending anomalies.

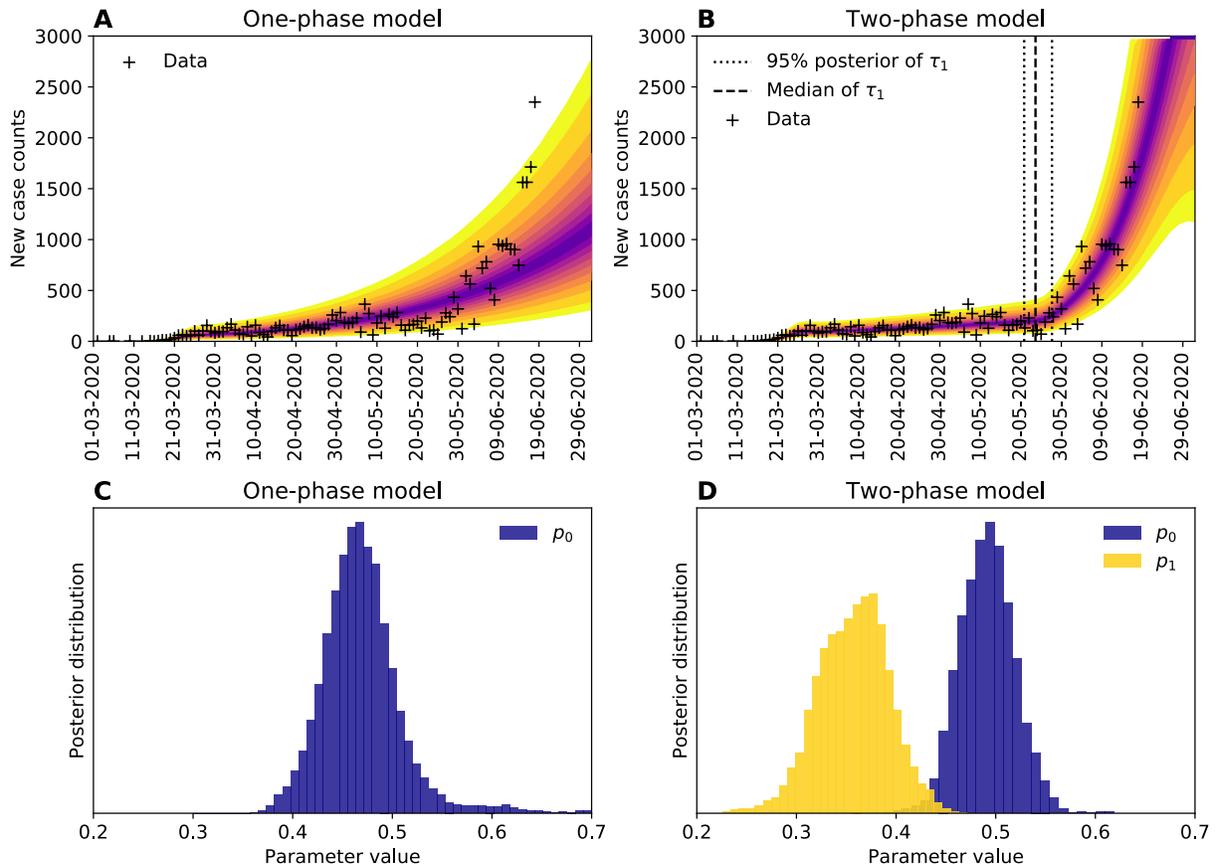

Figure 10. Predictions of the compartmental model **(A)** with consideration of only one period of social distancing ($n = 0$) and **(B)** with consideration of an initial period of social distancing followed by a distinct period of relatively lax adherence to social-distancing practices ($n = 1$) for the Phoenix metropolitan statistical area (MSA). In panel **(C)**, the marginal posterior for the social-distancing setpoint parameter $p_0$ inferred in the analysis of (A) is shown. In panel **(D)**, the marginal posteriors for the social-distancing parameters $p_0$ and $p_1$ inferred in the analysis of (B) are shown. Model selection indicates that the two-phase model is to be preferred (Appendix Table 1). In this analysis, we used data available from 21-January-2020 to 18-June-2020 (inclusive dates).

**Appendix Table 1.** Strength-of-evidence comparison of compartmental models accounting for an initial social-distancing period only ($n = 0$) and for an initial social-distancing period followed by a distinct second-phase social distancing period ($n = 1$).

| MSA | ΔAIC* | ΔBIC* | $p_0^{n=0}$ (95%)** | $p_0^{n=1}$ (95%)** | $p_1^{n=1}$ (95%)** |
|---|---|---|---|---|---|
| New York City | 17 | 8.6 | 0.88 (0.85–0.90) | 0.87 (0.80–0.89) | 0.36 (0.11–0.83) |
| Los Angeles | −6.5 | −15 | 0.45 (0.38–0.45) | 0.47 (0.42–0.80) | 0.38 (0.33–0.97) |
| Chicago | 18 | 9.5 | 0.57 (0.46–0.61) | 0.52 (0.46–0.75) | 0.25 (0.03–0.68) |
| Dallas | 18 | 9.4 | 0.52 (0.41–0.52) | 0.59 (0.49–0.77) | 0.41 (0.33–0.60) |
| Houston | 50 | 42 | 0.39 (0.34–0.45) | 0.49 (0.39–0.79) | 0.30 (0.20–0.56) |
| Washington | 1.0 | −7.5 | 0.39 (0.30–0.47) | 0.77 (0.71–0.80) | 0.68 (0.63–0.76) |
| Miami | 75 | 67 | 0.51 (0.46–0.57) | 0.92 (0.81–0.97) | 0.69 (0.61–0.80) |
| Philadelphia | 12 | 3.7 | 0.65 (0.57–0.69) | 0.55 (0.49–0.81) | 0.22 (0.03–0.69) |
| Atlanta | 9.9 | 1.5 | 0.54 (0.41–0.52) | 0.58 (0.44–0.78) | 0.29 (0.06–0.63) |
| Phoenix | 66 | 58 | 0.43 (0.37–0.49) | 0.55 (0.43–0.73) | 0.34 (0.26–0.54) |
| Boston | −31 | −39 | 0.36 (0.29–0.37) | 0.80 (0.69–0.85) | 0.18 (0.06–0.97) |
| San Francisco | 20 | 12 | 0.32 (0.29–0.35) | 0.36 (0.34–0.74) | 0.17 (0.07–0.63) |
| Riverside | 3.8 | −4.7 | 0.41 (0.36–0.46) | 0.43 (0.38–0.74) | 0.34 (0.03–0.48) |
| Detroit | 5.9 | −2.6 | 0.75 (0.60–0.78) | 0.80 (0.64–0.92) | 0.93 (0.14–0.97) |
| Seattle | 55 | 46 | 0.87 (0.75–0.90) | 0.82 (0.76–0.85) | 0.59 (0.48–0.68) |

*ΔAIC ≡ $AIC^{n=0} - AIC^{n=1}$ and ΔBIC ≡ $BIC^{n=0} - BIC^{n=1}$, where $AIC^{n=0}$ and $AIC^{n=1}$ are the Aikake information criterion (AIC) *(24)* values calculated for the $n = 0$ and $n = 1$ versions of the compartmental model and, similarly, $BIC^{n=0}$ and $BIC^{n=1}$ are the Bayesian information criterion (BIC) *(24)* values calculated for the $n = 0$ and $n = 1$ versions of the compartmental model. As recommended by Burnham and Anderson *(24)*, we interpret ΔAIC>10 to indicate "no support for $n = 0$" and ΔAIC<-10 to indicate "no support for $n = 1$." There are five MSAs for which ΔAIC and ΔBIC are both greater than 10: Houston, Miami, Phoenix, San Francisco, and Seattle. There is one MSA for which ΔAIC and ΔBIC are both less than −10: Boston. **The first entry in each row of this column is the maximum a posteriori (MAP) estimate. The next entry, a pair of numbers

within parentheses, indicates the 95% credible interval. In this analysis, we used data available from 21-January-2020 to 26-June-2020 (inclusive dates).

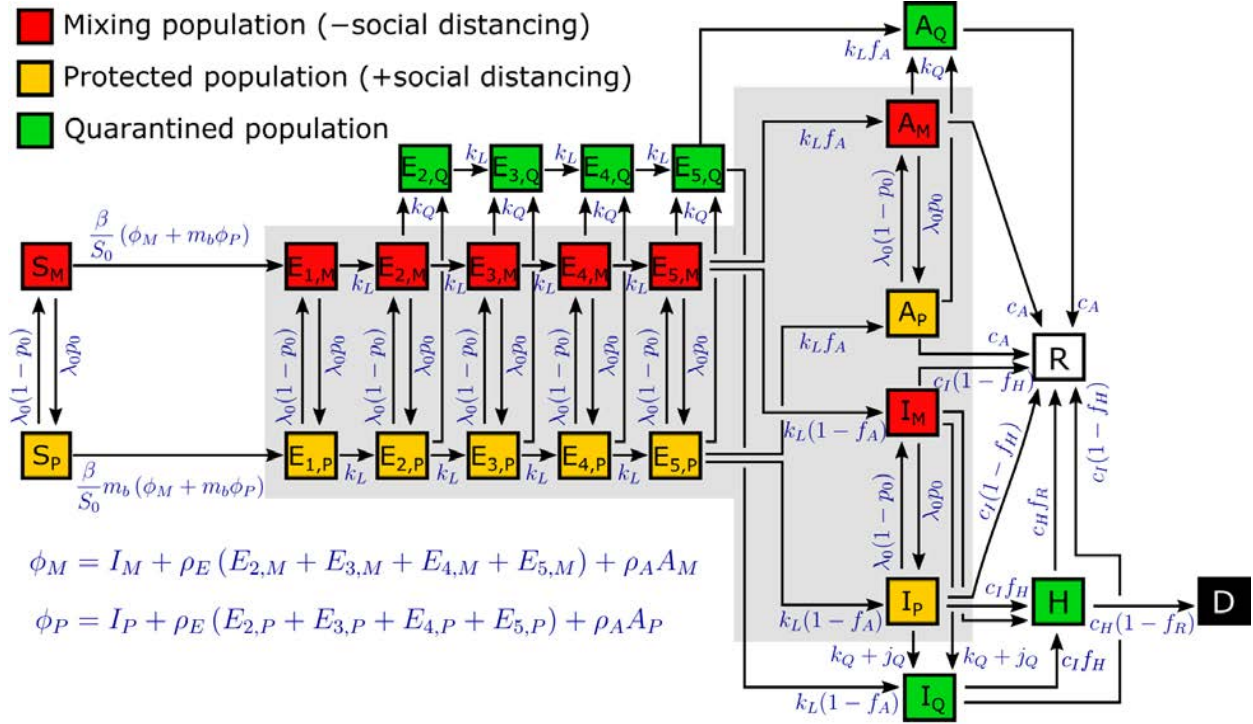

**Appendix Figure 1.** Detailed diagram of the populations and processes considered in the mechanistic compartmental model. In this illustration of the compartmental model, the labels attached to arrows indicate the parameters and variables that affect the rates of the processes represented by the arrows. There is a one-to-one correspondence between arrow labels and terms on the right-hand sides of Equations (5)–(21). The diagram is otherwise the same as that shown in Figure 2.

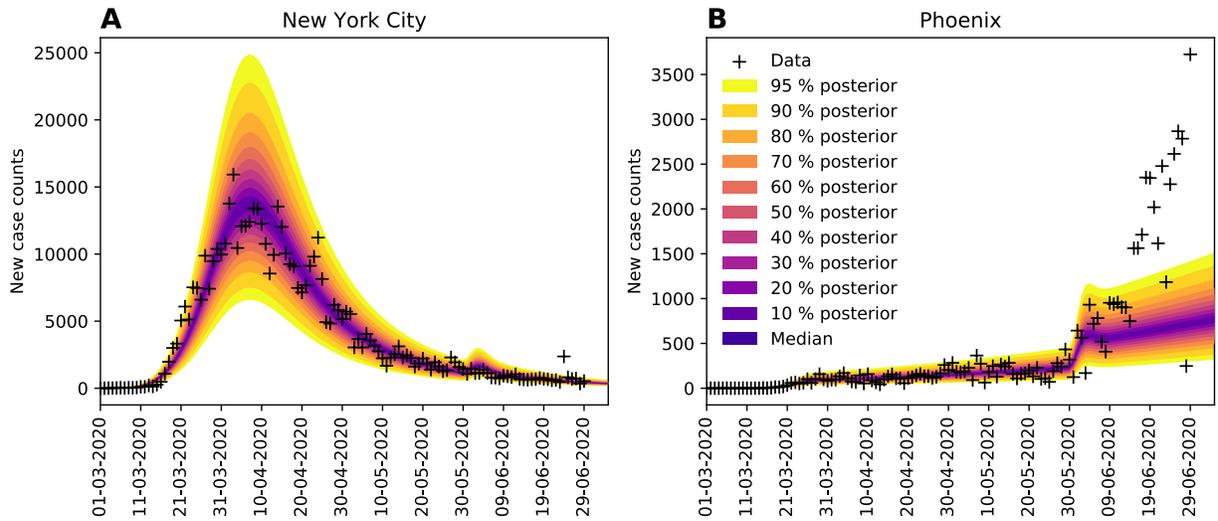

**Appendix Figure 2.** We evaluated the potential impact of a one-time mass gathering that causes 50,000 and 5,000 individuals to become newly infected on 30-May-2020 in **(A)** the New York City metropolitan statistical area (MSA) and **(B)** the Phoenix MSA, respectively. As can be seen, such an event will step the epidemic curve up without significantly changing the slope of the curve. According to the model, the trend in the slope of the curve is determined by the sustained level of adherence to effective social-distancing practices. We conclude that the recent trajectory of the epidemic curve of the Phoenix MSA cannot be explained by a one-time mass gathering.

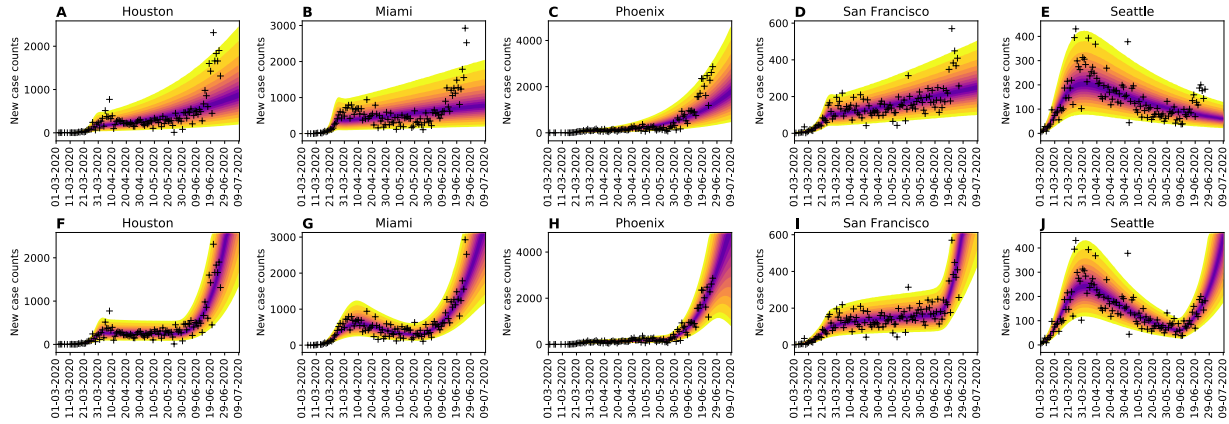

**Appendix Figure 3.** Predictive inferences conditioned on the compartmental model with either **(A)–(E)** one or **(F)–(J)** two distinct social-distancing periods. We consider five metropolitan statistical areas (MSAs), which according to Appendix Table 1, have epidemic curves better explained by the compartmental model with two social-distancing periods than by the compartmental model with just one. Maximum a posteriori (MAP) estimates for $\tau_1$ indicate that the second social distancing period began on 27-May-2020 for Houston, 19-April-2020 for Miami, 24-May-2020 for Phoenix, 12-June-2020 for San Francisco, and 07-June-2020 for Seattle. In this analysis, we used the data available from 21-January-2020 to 26-June-2020 (inclusive dates).

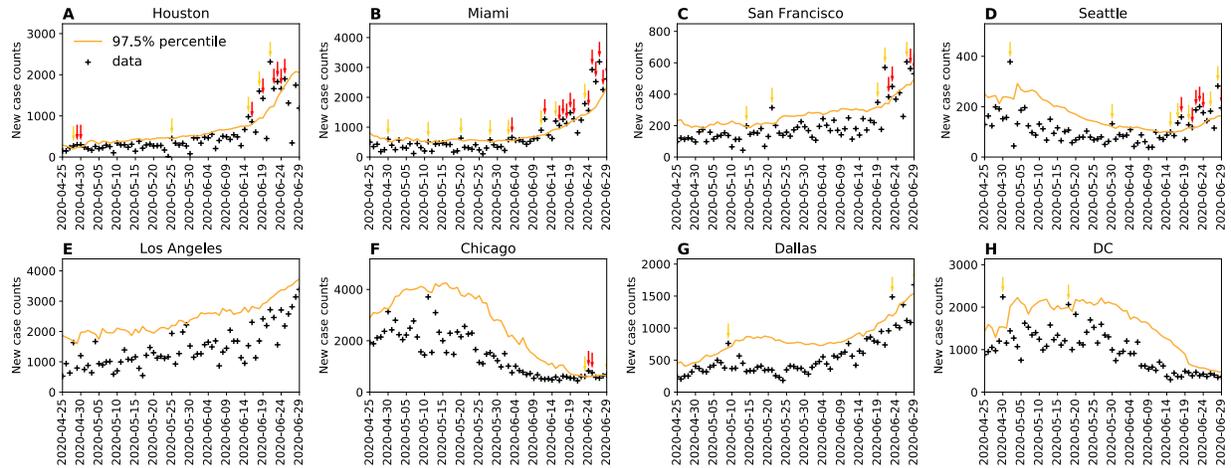

**Appendix Figure 4.** Comparison of next-day predictions and the corresponding empirical case reports for **(A)** Houston, **(B)** Miami, **(C)** San Francisco, **(D)** Seattle, **(E)** Los Angeles, **(F)** Chicago, **(G)** Dallas, and **(H)** Washington, DC. Like Phoenix, Houston, Miami, San Francisco, and Seattle have epidemic curves that are better explained by the compartmental model with two social-distancing periods vs. just one (Appendix Table 1). Upward-trending anomalies were detected for each of these MSAs. Los Angeles, Chicago, Dallas, and Washington, DC have epidemic curves better explained by the compartmental model with just one social-distancing period vs. two (Appendix Table 1). Upward-trending anomalies were not detected for three of these MSAs. The exception is Chicago. Two anomalies were detected for Chicago near the end of the period considered in the analysis.

**Video 1.** An animation showing daily predictive inferences made for the New York City metropolitan statistical area from 19-Mar-2020 to 6-Jun-2020 (inclusive dates). Inferences are conditioned on the single-phase ($n = 0$) compartmental model.

**Video 2.** An animation showing daily predictive inferences made for the Phoenix metropolitan statistical area from 30-Mar-2020 to 17-Jun-2020 (inclusive dates). Inferences are conditioned on the single-phase ($n = 0$) compartmental model.

**Video 3.** An animation showing daily predictive inferences made for the Houston metropolitan statistical area from 30-Mar-2020 to 17-Jun-2020 (inclusive dates). Inferences are conditioned on the single-phase ($n = 0$) compartmental model.

**Video 4.** An animation showing daily predictive inferences made for the Miami metropolitan statistical area from 30-Mar-2020 to 17-Jun-2020 (inclusive dates). Inferences are conditioned on the single-phase ($n = 0$) compartmental model.

**Video 5.** An animation showing daily predictive inferences made for the San Francisco metropolitan statistical area from 30-Mar-2020 to 17-Jun-2020 (inclusive dates). Inferences are conditioned on the single-phase ($n = 0$) compartmental model.

**Video 6.** An animation showing daily predictive inferences made for the Seattle metropolitan statistical area from 30-Mar-2020 to 17-Jun-2020 (inclusive dates). Inferences are conditioned on the single-phase ($n = 0$) compartmental model.

**Video 7.** An animation showing daily predictive inferences made for the Los Angeles metropolitan statistical area from 30-Mar-2020 to 17-Jun-2020 (inclusive dates). Inferences are conditioned on the single-phase ($n = 0$) compartmental model.

**Video 8.** An animation showing daily predictive inferences made for the Chicago metropolitan statistical area from 30-Mar-2020 to 17-Jun-2020 (inclusive dates). Inferences are conditioned on the single-phase ($n = 0$) compartmental model.

**Video 9.** An animation showing daily predictive inferences made for the Dallas metropolitan statistical area from 30-Mar-2020 to 17-Jun-2020 (inclusive dates). Inferences are conditioned on the single-phase ($n = 0$) compartmental model.

**Video 10.** An animation showing daily predictive inferences made for the Washington, DC metropolitan statistical area from 30-Mar-2020 to 17-Jun-2020 (inclusive dates). Inferences are conditioned on the single-phase ($n = 0$) compartmental model.

**Appendix Text**

*Full Description of the Curve-Fitting Model*

For each metropolitan statistical area (MSA) of interest, we assume that there is an infection curve $Q(t)$ describing the number of individuals who become infected at time $t$ with SARS-CoV-2 and who will later be detected in local COVID-19 surveillance efforts. Furthermore, we assume that this curve has a shape that can be generated/reproduced by $\rho_\Gamma(k, \theta, t)$, the probability density function (PDF) of a gamma distribution $\Gamma(k, \theta)$. In other words, we assume that $Q(t) = N\rho_\Gamma(k, \theta, t)$, where $N$ is a scaling factor that we can identify as the number of individuals who will be detected over the entire course of the local epidemic. The shape of a gamma distribution is flexible and determined by the values of its two parameters: $k$, which is called the shape parameter, and $\theta$, which is called the scale parameter. The functional form that we assume for $Q(t)$ allows the curve-fitting model to reproduce the shape of an epidemic curve having two timescales. Many empirical COVID-19 epidemic curves appear to have two timescales: an initial period during which new case reports increase relatively quickly from day to day followed by a period during which new case reports decrease relatively slowly from day to day.

We do not take the infection curve $Q(t)$ to correspond directly to the number of new COVID-19 cases reported on the date encompassing time $t$, because only symptomatic individuals are likely to be detected in COVID-19 surveillance testing (to a first approximation). This situation complicates our model as there is known to be a potentially lengthy, variable delay in the onset of symptoms after infection *(12)*. We assume that the waiting time $\tau - t$ for the onset of COVID-19 symptoms after SARS-CoV-2 infection at time $t$ is distributed according to a log-normal distribution. Let us use $\rho_{LN}(\tau - t; \mu, \sigma)$ to denote the PDF of the waiting-time distribution modeled by a log-normal distribution with parameters $\mu$ and $\sigma$ set to the values estimated by Lauer et al. *(12)*. Let us use $I(t_i, t_{i+1})$ to denote the predicted number of new COVID-19 cases reported within a period beginning at time $t_i \equiv t_0 + i$ d and ending at time $t_{i+1}$, where $t_0 > 0$ is the start time of the local epidemic. We assume that surveillance testing for SARS-CoV-2 infection starts prior to time $t_0$, and we take time $t = 0$ to correspond to 0000 hours on 21-January-2020, the date on which detection of the first US COVID-19 case was widely reported *(3)*. Under the aforementioned assumptions, $I(t_i, t_{i+1})$ is given by a convolution of integral functions. Namely, $I(t_i, t_{i+1})$ is given by the following expression:

$$I(t_i, t_{i+1}) = N \int_{t_i}^{t_{i+1}} \int_{t_0}^{\tau} \rho_{LN}(\tau - s; \mu, \sigma)\, \rho_\Gamma(s - t_0; k, \theta)\, \mathrm{d}s\, \mathrm{d}\tau \quad (1)$$

It should be noted that $s$ in this expression is a dummy variable of integration.

Equation (1) can be evaluated through numerical quadrature, but this procedure is computationally expensive. To overcome this limitation, we replace the double integral in

Equation (1) with a sum, and we calculate $I(t_i, t_{i+1})$ using the following expression instead of Equation (1):

$$I(t_i, t_{i+1}) = K_0 Q_i + K_1 Q_{i-1} + \cdots + K_{i-1} Q_1 + K_i Q_0 = \sum_{j=0}^{i} K_{i-j} Q_j \qquad (2)$$

where

$$K_{i-j} = \int_{t_{i-j}}^{t_{i-j+1}} \rho_{LN}(t; \mu, \sigma) dt = F_{LN}(t_{i-j+1}; \mu, \sigma) - F_{LN}(t_{i-j}; \mu, \sigma) \qquad (3)$$

and

$$Q_j = N \int_{t_j}^{t_{j+1}} \rho_\Gamma(t - t_0; k, \theta) dt = N[F_\Gamma(t_{j+1} - t_0; k, \theta) - F_\Gamma(t_j - t_0; k, \theta)] \qquad (4)$$

In Equation (2), the $K_{i-j}$ terms are weighting functions (i.e., kernels) that account for the variable duration of the incubation period, and the $Q_j$ terms represent cumulative numbers of new detectable infections occurring over discrete 1-d periods. In Equation (3), each $F_{LN}$ term denotes a cumulative distribution function (CDF) of a log-normal distribution, and in Equation (4), each $F_\Gamma$ term denotes a CDF of a gamma distribution. In other words, $Q_j$ is the cumulative number of individuals infected in the period $(t_j, t_{j+1})$ who will eventually be detected, and $K_{i-j}$ is the probability that one of these individuals becomes symptomatic and is detected in the period $(t_i, t_{i+1})$, where $t_i \geq t_j$.

The functional form of our curve-fitting model is defined by Equations (2)–(4), which are derived from Equation (1). As can be seen by inspecting Equation (1), the curve-fitting model has six parameters: $N$, $t_0$ (which is hidden in the definition of $t_i$), $k$, $\theta$, $\mu$, and $\sigma$. As noted earlier, estimates are available for $\mu$ and $\sigma$ from Lauer et al. *(12)*. These parameters characterize the variable duration of the incubation period, which starts at infection and ends at the onset of symptoms. Thus, we take $\mu$ and $\sigma$ to have fixed region-independent values. We take the remaining parameters—$N$ (a population size/scaling factor), $t_0$ (the start time of the local epidemic), and $k$ and $\theta$ (the parameters that determine the shape of the infection curve $Q(t)$)—to have adjustable region-specific values. In our daily inferences, we consider one additional region-specific adjustable parameter, the dispersal parameter of the likelihood function (see Equation (31) below). The value of this parameter, $r$, is inferred jointly with the values of $N$, $t_0$, $k$, and $\theta$.

*Full Description of the Mechanistic Compartmental Model*

The compartmental model, which is illustrated in Appendix Figure 1, consists of the following 25 ordinary differential equations (ODEs):

$$\frac{dS_M}{dt} = -\beta\left(\frac{S_M}{S_0}\right)(\phi_M(t,\rho) + m_b\phi_P(t,\rho)) - U_\sigma(t)\Lambda_\tau(t)[P_\tau(t)S_M - (1 - P_\tau(t))S_P] \quad (5)$$

$$\frac{dS_P}{dt} = -m_b\beta\left(\frac{S_P}{S_0}\right)(\phi_M(t,\rho) + m_b\phi_P(t,\rho))$$
$$+ U_\sigma(t)\Lambda_\tau(t)[P_\tau(t)S_M - (1 - P_\tau(t))S_P] \quad (6)$$

$$\frac{dE_{1,M}}{dt} = \beta\left(\frac{S_M}{S_0}\right)(\phi_M(t,\rho) + m_b\phi_P(t,\rho)) - k_L E_{1,M}$$
$$- U_\sigma(t)\Lambda_\tau(t)[P_\tau(t)E_{1,M} - (1 - P_\tau(t))E_{1,P}] \quad (7)$$

$$\frac{dE_{1,P}}{dt} = m_b\beta\left(\frac{S_P}{S_0}\right)(\phi_M(t,\rho) + m_b\phi_P(t,\rho)) - k_L E_{1,P}$$
$$+ U_\sigma(t)\Lambda_\tau(t)[P_\tau(t)E_{1,M} - (1 - P_\tau(t))E_{1,P}] \quad (8)$$

$$\frac{dE_{i,M}}{dt} = k_L E_{i-1,M} - k_L E_{i,M} - k_Q E_{i,M} - U_\sigma(t)\Lambda_\tau(t)[P_\tau(t)E_{i,M} - (1 - P_\tau(t))E_{i,P}],$$
$$\text{for } i = 2, 3, 4, 5 \quad (9)$$

$$\frac{dE_{i,P}}{dt} = k_L E_{i-1,P} - k_L E_{i,P} - k_Q E_{i,P} + U_\sigma(t)\Lambda_\tau(t)[P_\tau(t)E_{i,M} - (1 - P_\tau(t))E_{i,P}],$$
$$\text{for } i = 2, 3, 4, 5 \quad (10)$$

$$\frac{dE_{2,Q}}{dt} = k_Q(E_{2,M} + E_{2,P}) - k_L E_{2,Q} \quad (11)$$

$$\frac{dE_{i,Q}}{dt} = k_Q(E_{i,M} + E_{i,P}) + k_L E_{i-1,Q} - k_L E_{i,Q}, \text{ for } i = 3, 4, 5 \quad (12)$$

$$\frac{dA_M}{dt} = f_A k_L E_{5,M} - k_Q A_M - U_\sigma(t)\Lambda_\tau(t)[P_\tau(t)A_M - (1 - P_\tau(t))A_P] - c_A A_M \quad (13)$$

$$\frac{dA_P}{dt} = f_A k_L E_{5,P} - k_Q A_P + U_\sigma(t)\Lambda_\tau(t)[P_\tau(t)A_M - (1 - P_\tau(t))A_P] - c_A A_P \quad (14)$$

$$\frac{dA_Q}{dt} = f_A k_L E_{5,Q} + k_Q(A_M + A_P) - c_A A_Q \quad (15)$$

$$\frac{dI_M}{dt} = (1 - f_A)k_L E_{5,M} - (k_Q + j_Q)I_M - U_\sigma(t)\Lambda_\tau(t)[P_\tau(t)I_M - (1 - P_\tau(t))I_P]$$
$$- c_I I_M \quad (16)$$

$$\frac{dI_P}{dt} = (1 - f_A)k_L E_{5,P} - (k_Q + j_Q)I_P + U_\sigma(t)\Lambda_\tau(t)[P_\tau(t)I_M - (1 - P_\tau(t))I_P] - c_I I_P \quad (17)$$

$$\frac{dI_Q}{dt} = (1 - f_A)k_L E_{5,Q} + (k_Q + j_Q)(I_M + I_P) - c_I I_Q \tag{18}$$

$$\frac{dH}{dt} = f_H c_I (I_M + I_P + I_Q) - c_H H \tag{19}$$

$$\frac{dD}{dt} = (1 - f_R) c_H H \tag{20}$$

$$\frac{dR}{dt} = c_A (A_M + A_P + A_Q) + (1 - f_H) c_I (I_M + I_P + I_Q) + f_R c_H H \tag{21}$$

where $\beta, S_0, m_b, k_L, k_Q, j_Q, f_A, f_H, f_R, c_A, c_I$, and $c_H$ are positive-valued time-invariant parameters. (It should be noted that parameter names are unique but only within the namespace of a given model.) Each ODE in Equations (5)–(21) defines the time-rate of change of a (sub)population, i.e., the time-rate of change of a state variable. There are 25 state variables, one for each ODE. Note that Equations (9), (10), and (12) define 4, 4, and 3 ODEs of the model, respectively.

The initial condition is taken to be $S_M(t_0) = S_0$, $I_M(t_0) = I_0 = 1$, with all other populations $(S_P, E_{1,M}, \ldots, E_{5,M}, E_{1,P}, \ldots, E_{5,P}, E_{2,Q}, \ldots, E_{5,Q}, A_M, A_P, A_Q, I_P, I_Q, H, D,$ and $R)$ equal to 0. The parameter $S_0$ denotes the total region-specific population size. Thus, we assume that the entire population is susceptible at the start of the epidemic at time $t = t_0 > 0$, where time $t = 0$ is 0000 hours on 21-January-2020. The parameter $I_0$, which we always take to be 1, denotes the number of infectious symptomatic individuals at the start of the regional epidemic.

Subscripts attached to state variables are used to denote subpopulations. The subscripts $M$ and $P$ are attached to variables representing mixing and protected populations, respectively. For example, the variables $S_M$ and $S_P$ denote the population sizes of mixing and protected individuals who are susceptible to infection. Individuals in a protected population practice social distancing; individuals in a mixing population do not. The approach that we have taken to model social distancing is similar to that of Anderson et al. *(30)*.

The incubation period is divided into five stages. The numerical subscripts 1, 2, 3, 4, and 5 attached to $E$ variables indicate progression through these five stages. Exposed individuals in the incubation period, except for those in the first stage, are taken to be infectious. They are also taken to lack symptoms. They are either presymptomatic (i.e., individuals who will later develop symptoms) or asymptomatic (i.e., individuals who will never develop symptoms).

The subscript $Q$ is attached to variables representing populations of quarantined individuals. The state variable $I_Q$ is a special case; it accounts for symptomatic individuals who are quarantined as well as individuals who are self-isolating because of symptom awareness.

The parameter $k_Q$ characterizes the rate at which infected individuals move into quarantine because of testing and contact tracing. The parameter $j_Q$ characterizes the rate at which symptomatic individuals self-isolate because of symptom awareness. We recognize that susceptible individuals may enter quarantine (through contact tracing) but we assume that the size of the quarantined population is negligible compared to that of the total susceptible population and that susceptible individuals entering quarantine leave quarantine as susceptible individuals.

The parameters $\beta$ and $m_b < 1$ characterize transmission of disease: $\beta$ characterizes the rate of transmission attributable to contacts between two mixing individuals, $m_b \beta$ characterizes the rate of transmission attributable to contacts between one mixing and one protected individual, and $m_b^2 \beta$ characterizes the rate of transmission attributable to contacts between two protected individuals. Infectious individuals taken to contribute to COVID-19 transmission include those in the following pools: $E_{2,M}, \ldots, E_{5,M}$ and $E_{2,P}, \ldots, E_{5,P}$, $A_M$ and $A_P$, and $I_M$ and $I_P$. Recall that we do not consider individuals in the first stage of the incubation period (i.e., individuals in $E_1$ pools) to be infectious. The assumption is that these individuals are not shedding enough virus to be infectious (or detectable in surveillance testing).

The variables $E_{1,M}, \ldots, E_{5,M}$ and $E_{1,P}, \ldots, E_{5,P}$ denote the population sizes of mixing and protected exposed individuals in the five stages of the incubation period. The variables $E_{2,Q}, \ldots, E_{5,Q}$ denote the population sizes of quarantined exposed individuals in the five stages. There is no $E_{1,Q}$ population, as we assume that individuals in the first stage of the incubation period are unlikely to test positive for SARS-CoV-2 or to be reached in contact tracing efforts before leaving the $E_1$ state. The parameter $k_L$ characterizes disease progression, from one stage of the incubation period to the next and ultimately to an immune clearance phase. Individuals leaving the $E_5$ pools enter the immune clearance phase, meaning that they become eligible for recovery. An individual leaving an $E_5$ pool with symptom onset enters an $I$ pool, whereas an individual leaving an $E_5$ pool without symptom onset enters an $A$ pool. Individuals in $I$ pools are considered to have mild disease (with the possibility to progress to severe disease).

The dynamics of social distancing are characterized by three step functions (i.e., piecewise constant functions having only finitely many pieces): $U_\sigma$, $\Lambda_\tau$, and $P_\tau$. The subscripts attached to these functions denote times: $\sigma$ is a particular time, whereas $\tau$ is a set of times, as discussed later. The value of $U_\sigma$ switches from 0 to 1 at time $t = \sigma > t_0$, the start of an initial social-distancing period. As discussed later, the function $\Lambda_\tau$ defines a timescale for change in social-distancing practices for one or more distinct periods of social distancing, and the function $P_\tau$ establishes a setpoint for the fraction of the total population of susceptible and infectious, non-quarantined/non-self-isolated/non-hospitalized individuals adhering to social-distancing practices for one or more distinct periods of social distancing.

The parameter $f_A$ denotes the fraction of infected individuals who never develop symptoms (i.e., the fraction of all cases that are asymptomatic). The variables $A_M$ and $A_P$ denote the sizes of the populations of mixing and protected individuals who have been infected, progressed through the incubation period, are currently in the immune clearance phase, and will never develop symptoms. The parameter $c_A$ characterizes the rate at which asymptomatic individuals recover.

The variable $R$ tracks recoveries of asymptomatic individuals, symptomatic individuals with mild disease, and symptomatic (hospitalized) individuals with severe disease. All individuals who recover are assumed to have immunity.

The variables $I_M$ and $I_P$ denote the sizes of the populations of mixing and protected symptomatic individuals with mild disease. The parameter $c_I$ characterizes the rate at which symptomatic individuals with mild disease recover or progress to severe disease. The parameter $f_H$ is the fraction of symptomatic individuals who progress to severe disease requiring hospitalization. As a simplification, we assume that all individuals with severe disease are hospitalized. The variable $H$ represents the population size of hospitalized individuals, which are taken to be quarantined. Thus, the model does not consider nosocomial transmission.

The parameter $f_R$ denotes the fraction of (hospitalized) individuals with severe disease who recover. The parameter $c_H$ characterizes the hospital discharge rate, i.e., the rate at which hospitalized individuals with severe disease either recover or die. The variable $D$ tracks deaths.

The time-dependent terms $\phi_M(t,\rho)$ and $\phi_P(t,\rho)$ appearing in Equations (5)–(8) represent the effective population sizes of infectious individuals in the mixing and protected subpopulations, respectively. These quantities are defined as follows:

$$\phi_M(t,\rho) \equiv I_M + \rho_E(E_{2,M} + E_{3,M} + E_{4,M} + E_{5,M}) + \rho_A A_M \qquad (22)$$

$$\phi_P(t,\rho) \equiv I_P + \rho_E(E_{2,P} + E_{3,P} + E_{4,P} + E_{5,P}) + \rho_A A_P \qquad (23)$$

where $\rho = (\rho_E, \rho_A)$, $\rho_E$ is a constant characterizing the relative infectiousness of presymptomatic individuals compared to symptomatic individuals and $\rho_A$ is a constant characterizing the relative infectiousness of asymptomatic individuals compared to symptomatic individuals. Recall that we assume that individuals in the first stage of the incubation period (i.e., individuals in either the $E_{1,M}$ or $E_{1,P}$ population) are not infectious. We also assume that the individuals in these populations cannot be quarantined until after transitioning to the $E_{2,M}$ or $E_{2,P}$ population (because they are assumed to test negative and because contact tracing is assumed to be too slow to catch individuals in the transient first stage of incubation). Recall that individuals in the $A_M$, $A_P$, and $A_Q$ populations are defined as individuals who became infected, passed through all five stages of the incubation period, are currently in the immune clearance phase, and will never develop symptoms. Thus, individuals in the exposed $E$ populations include both presymptomatic individuals (i.e., individuals who will enter the $I$ populations) and asymptomatic individuals (i.e., individuals who will enter the $A$ populations).

The time-dependent terms $U_\sigma(t)$, $P_\tau(t)$, and $\Lambda_\tau(t)$ appearing in Equations (5)–(10), Equations (13) and (14) and Equations (16) and (17) are step functions defined as follows:

$$U_\sigma(t) = \begin{cases} 0 & t < \sigma \\ 1 & t \geq \sigma \end{cases} \quad (24)$$

$$P_\tau(t) = \begin{cases} p_0 & \sigma \leq t < \tau_1 \\ p_1 & \tau_1 \leq t < \tau_2 \\ \vdots & \vdots \\ p_n & \tau_n \leq t < \infty \end{cases} \quad (25)$$

$$\Lambda_\tau(t) = \begin{cases} \lambda_0 & \sigma \leq t < \tau_1 \\ \lambda_1 & \tau_1 \leq t < \tau_2 \\ \vdots & \vdots \\ \lambda_n & \tau_n \leq t < \infty \end{cases} \quad (26)$$

where $\sigma > t_0$ is the time at which widespread social distancing initially begins, the integer $n \geq 0$ is the number of societal (major/widespread) shifts in social-distancing practices after the initial onset of social distancing, each $p_i < 1$ is a parameter characterizing the quasi-stationary fraction of susceptible individuals practicing social distancing during the $(i+1)$th period of social distancing, each $\lambda_i$ is a constant defining a timescale for change in social-distancing practices during the $(i+1)$th period of social distancing, $\tau = \{\tau_0, \ldots, \tau_{n+1}\}$, $\tau_0 \equiv \sigma$, $\tau_{n+1} \equiv \infty$, and $\tau_{i+1} > \tau_i$ for $i = 0, \ldots, n-1$. The value of $P_\tau(t)$ defines a setpoint for the quasi-stationary size of the protected population of susceptible individuals: $P_\tau(t) \times 100\%$ of the total susceptible population. The value of $\Lambda_\tau(t)$ determines how quickly the setpoint is reached. As indicated in Equations (25) and (26), we only consider step-changes in the values of $P_\tau(t)$ and $\Lambda_\tau(t)$, a simplification. Thus, for a period during which social-distancing practices are intensifying (relaxing), we increase (decrease) the value of $P_\tau(t)$ at the start of the period in a step-change and then hold it constant until the next step-change, if any.

*Full Description of the Auxiliary Measurement Model*

To determine how consistent a particular parameterization of the compartmental model is with available COVID-19 surveillance data, we need to define a quantity—a model output—that corresponds to daily reports of the number of new confirmed COVID-19 cases. Case reporting by public health officials is typically daily. We expect that the vast majority of cases are detected because of symptom-driven (vs. random) testing and/or presentation in a clinical setting. Accordingly, as a simplification, we assume that individuals detected in surveillance are symptomatic. To define a model output comparable to the number of new cases reported on a given day, we start by considering the predicted cumulative number of presymptomatic individuals who become symptomatic while evading quarantine (because of contact tracing) until at least the onset of symptoms, which we will denote as $C_S$. According to the model, the time rate of change of $C_S$ is given by the following equation:

$$\frac{dC_S}{dt} = (1 - f_A)k_L(E_{5,M} + E_{5,P}) \tag{27}$$

The right-hand side of this equation gives the rate at which non-quarantined presymptomatic individuals exit the incubation period and enter the immune clearance phase, in which they are symptomatic and therefore taken to be detectable in local surveillance efforts.

Equation (27) and the ODEs of the compartmental model form a coupled system of equations, which can be numerically integrated to obtain trajectories for the state variables and $C_S$, the expected cumulative number of symptomatic cases. From the trajectory for $C_S$, we obtain a prediction for $I(t_i, t_{i+1})$, the expected number of new COVID-19 cases reported on a given calendar date $\mathcal{D}_i$, from the following equation:

$$I(t_i, t_{i+1}) = f_D[C_S(t_{i+1}) - C_S(t_i)] \tag{28}$$

where $f_D$ is taken to be an adjustable region-specific parameter characterizing the time-averaged fraction of symptomatic cases detected. Equation (28) completes the formulation of our measurement model. $I(t_i, t_{i+1})$ is the model output that we compare to $\delta C_i$, the number of new cases reported on calendar date $\mathcal{D}_i$.

*The Adjustable and Fixed Parameters of the Compartmental Model and Auxiliary Measurement Model*

The parameters of the compartmental model (Equations (5)–(26)) and the auxiliary measurement model (Equations (27) and (28)) are taken to have either adjustable or fixed values. The adjustable parameter values are estimated (daily) through Bayesian inference on the basis of surveillance data (i.e., reports of newly detected cases). The fixed parameter values are held constant during inference; they are based on non-surveillance data and/or assumptions, which are discussed in the section below. In this section, we simply delineate the parameters with adjustable and fixed values. The compartmental model formulated for a given regional epidemic has a total of $16 + 3(n + 1)$ parameters. The value of $n$ is structural; it sets the number social-distancing periods considered.

The value of $n$ corresponds to the number of periods of distinct social-distancing behaviors that follow an initial period of social distancing, which we take to begin at time $t = \sigma > t_0$. Here, we take $n = 0$ or 1 for all regional epidemics of interest. Usually, we set $n = 0$. In cases where we set $n = 1$, this setting was motivated by second wave-type dynamics suggested by the surveillance data, which we take to indicate a relaxation of social-distancing practices at time $t = \tau_1 > \sigma$. The parameters of the initial social-distancing period are $\sigma$, $p_0$, and $\lambda_0$. The parameters of the second social-distancing period, if considered, are $\tau_1$, $p_1$, and $\lambda_1$. Thus, there are $3(n + 1)$ social-distancing parameters, all of which are taken to be adjustable.

In addition to the $3(n + 1)$ social-distancing parameters, we have 16 other parameters. Three of these define the initial condition: $t_0$, $S_M(t = t_0) = S_0$, and $I_M(t = t_0) = I_0$, where $t_0$ is the time at which the epidemic begins, $S_0$ is taken to be the total population of the region of interest, and $I_0$ (the initial number of infected individuals) is always taken to be 1 (an assumption). We take $t_0$ to be adjustable and $S_0$ and $I_0$ to be fixed. The value of $S_0$ is set on the basis of population estimates by the US Census Bureau for the metropolitan statistical areas of interest *(13)*, which are delineated by the US Office of Management and Budget *(10)*.

There is only one more adjustable parameter of the compartmental model: $\beta$, which characterizes the rate of disease transmission attributable to contacts among individuals within the mixing population. In the period before the onset of social distancing, from $t_0$ to $\sigma$, when $S_M/S_0 \approx 1$, the instantaneous rate of disease transmission is $\beta \phi_M(t, \rho)$, where $\phi_M(t, \rho)$ is the effective number of infectious individuals at time $t$, a weighted sum of the numbers of symptomatic, presymptomatic, and asymptomatic individuals determined by $\rho = (\rho_E, \rho_A)$. We assume that exposed individuals after the first stage of disease incubation are infectious, as are asymptomatic individuals in the immune clearance phase who have passed through all 5 stages of disease incubation and who will never develop symptoms.

The remaining 12 parameters of the compartmental model, which are taken to have fixed, region-independent values, are as follows: $m_b$, $\rho_E$, $\rho_A$, $k_L$, $k_Q$, $j_Q$, $f_A$, $f_H$, $c_A$, $c_I$, $f_R$ and $c_H$. Our estimates for these parameters are discussed in the section immediately below. It should be noted that settings for $f_R$ and $c_H$ do not affect predictions of new cases because these parameters characterize recovery/morbidity of hospitalized individuals. The parameter $f_R$ is the fraction of hospitalized individuals who recover, and the parameter $c_H$ characterizes the hospital discharge rate. Although nosocomial disease transmission is a significant concern, we assume that hospitalized individuals are effectively quarantined such that the overall rate of disease transmission in a given region is insensitive to the number of hospitalized individuals in that region.

*Estimates of 12 Fixed Parameter Values of the Compartmental Model*

Here, we summarize the rationale/justification for each of our estimates for the values of the following 12 parameters of the compartmental model: $m_b$, $\rho_E$, $\rho_A$, $k_L$, $k_Q$, $j_Q$, $f_A$, $f_H$, $c_A$, $c_I$, $f_R$, and $c_H$. These estimates are assumed to apply to all regions, i.e., we take these parameters to have region-independent values.

The parameter $m_b$ characterizes the effects of social distancing on disease transmission. Without social distancing, all contacts responsible for disease transmission are between mixing individuals (i.e., between individuals in the $I_M$ and $S_M$ pools) and the rate of transmission is characterized by $\beta$. With social distancing, there are contacts involving 1 individual in a mixing population and 1 individual in a protected population (e.g., between individuals in the $I_M$ and $S_P$ pools or in the $S_M$ and $I_P$ pools) and also contacts involving 2 individuals in protected populations (e.g., between individuals in the $I_P$ and $S_P$ pools). In the model, the rates of transmission associated with these types of contacts are characterized by $m_b\beta$ and $m_b^2\beta$, respectively. We can be confident that social distancing is protective (i.e., $m_b < 1$) but there is little information available to suggest the magnitude of the effect. We arbitrarily set $m_b = 0.1$, which can be interpreted to mean that a susceptible individual practicing social distancing has a 10-fold smaller chance of becoming infected than a susceptible individual that is not practicing social distancing. In exploratory analyses, wherein we allowed $m_b$ to be a free parameter, we found that its inferred value is positively correlated with the extent of social distancing, which is determined by the relevant social-distancing setpoint parameter (viz., $p_0$). Thus, we interpret the inferred quasi-stationary value of $S_P$ to be an effective population size. If our estimate for $m_b$ is too high (i.e., we underestimate the protective effect of social distancing), the effective size will be larger than the true size. Conversely, if our estimate for $m_b$ is too low, the effective size will be smaller than the true size.

The parameters $\rho_E$ and $\rho_A$ characterize the relative infectiousness of individuals without symptoms during the incubation period and the immune clearance phase, respectively. Infectiousness is compared to that of a symptomatic individual. Using a one-step real-time reverse transcriptase-polymerase chain reaction (rRT-PCR) assay to quantify viral RNA abundance in nasopharyngeal and oropharyngeal samples, Arons et al. *(14)* determined rRT-PCR threshold cycle (Ct) values for 17 symptomatic and 24 presymptomatic individuals. The former group of individuals had typical symptoms, and the latter group of individuals lacked symptoms at the time of testing but later developed symptoms (within 1 week after testing). At the time of testing, the median Ct values for symptomatic and presymptomatic individuals were 24.8 and 23.1, respectively. (NB: Ct value is inversely proportional to abundance.) Based on these results and an assumption that infectiousness is proportional to viral load, we estimate that $\rho_E = 1.1$. An estimate for $\rho_E$ greater than 1 is consistent with the findings of He et al. *(25)*, who inferred that viral load is maximal 0.7 d before the onset of symptoms from an analysis of temporal viral load data and information available about infector-infectee transmission pairs. Over a period of 19 d, Nguyen et al. *(15)* performed daily rRT-PCR assays for viral RNA in nasopharyngeal samples from 17 symptomatic and 13 asymptomatic individuals. Ngyuen et al. *(15)* developed a curve-fitting model for each group to characterize their viral decay kinetics. These models indicate that the mean Ct value for symptomatic individuals was roughly 90% of the mean Ct value for asymptomatic individuals over the first week of the study, after which most individuals tested negative or had a Ct value near the threshold of detection, 40. Thus, we estimate that $\rho_A = 0.9$.

The parameter $k_L$ characterizes the duration of the incubation period. In the model, the incubation period is divided into 5 stages (for reasons explained shortly). The waiting time for completion of all 5 stages is described by an Erlang distribution with a shape parameter $k = 5$ and a scale parameter $\mu = 1/k_L$. Lauer et al. *(12)* estimated times of exposure and symptom onset for 181 confirmed cases and found that the median time between SARS-CoV-2 infection and onset of COVID-19 symptoms is 5.1 d. Lauer et al. *(12)* also found that the empirical distribution of waiting times is fit by an Erlang distribution with $k = 6$ and $\mu = 0.88$ d. This latter finding suggests that the empirical waiting time distribution can be reproduced by dividing the incubation period into 6 stages and setting $k_L = 1.14$ d$^{-1}$. However, an Erlang distribution with $k = 5$ and $\mu = 1.06$ d has a nearly identical shape. Because simulation costs are reduced by dividing the incubation period into 5 instead of 6 stages, we considered 5 stages in the model. The distribution of waiting times estimated by Lauer et al. *(12)* is reproduced by our model when we set $k_L = 0.94$ d$^{-1}$.

The parameters $k_Q$ and $j_Q$ characterize testing-driven quarantine and symptom-driven self-isolation. We assume that testing is random. Thus, the number of infected individuals moving into quarantine per d is the number of infected individuals subject to quarantine times the fraction of the total population tested per d times a multiplier capturing the effect of contact tracing. We take the multiplier to be average household size, 2.5 (US Census Bureau). Thus, based on approximately 500,000 tests per d in the US (https://covidtracking.com/data/us-daily) and a total population of 330 million (https://www.census.gov/popclock/), we estimate $k_Q = 0.0038$ d$^{-1}$. We assume $j_Q = 0.4$ d$^{-1}$. With this setting, the median waiting time from onset of symptoms to initiation of self-isolation is approximately 40 h. A faster timescale for self-isolation is probably not realistic despite general awareness of the COVID-19 pandemic, because as considered in the study of Böhmer et al. *(26)*, for any given individual, there may be a prodromal phase of ~1 d marked by non-COVID-19-specific symptoms other than fever and cough.

The parameter $f_A$ is the fraction of infected individuals who never develop symptoms. We estimate $f_A$ on the basis of information about the COVID-19 outbreak on the *Diamond Princess* cruise ship, as recounted by Sakurai et al. *(17)*. Before disembarking, 3,618 passengers and crew members were tested for SARS-CoV-2 infection. 410 of 712 individuals testing positive for SARS-CoV-2 were without symptoms at the time of testing. The Ministry of Health, Labour, and Welfare of Japan *(16)* reported that 311 (76%) of these individuals remained asymptomatic over the course of long-term follow-up. Thus, we estimate that $f_A = \frac{311}{712} \approx 0.44$. This estimate is consistent with the results of other studies. Lavezzo et al. *(27)* estimated that 43% of all infections are asymptomatic. In the study of Gudbjartsson et al. *(28)*, 7 of 13 individuals detected to have SARS-CoV-2 infection in random-sample population screening did not report symptoms; 43% of all SARS-CoV-2-positive participants in the study were symptom-free.

The parameter $f_H$ is the fraction of symptomatic individuals progressing to severe disease. We set $f_H$ such that our model predicts a uniform infection fatality rate (IFR) consistent with that determined by Perez-Saez et al. *(18)* from serological survey results and death incidence reports: 0.0064 (~0.64%). For a discussion of other IFR estimates, which tend to be similar, see Grewelle and De Leo *(29)*. According to our model, IFR is given by $(1 - f_A)f_H(1 - f_R)$. This quantity is the fraction of all infected individuals predicted to develop symptoms and then to progress to severe disease (and hospitalization) and finally a fatal outcome. Thus, based on our estimates for $f_A$ (0.44) and $f_R$ (0.79) and the empirical IFR (0.0064), we set $f_H = \frac{0.0064}{0.56 \times 0.21} \approx 0.054$.

The parameter $c_A$ characterizes the duration of infectiousness of asymptomatic individuals in the immune response phase. For each of 89 asymptomatic individuals, Sakurai et al. *(17)* reported the time between the first positive PCR test for SARS-CoV-2 and the first of two serial negative PCR tests. The mean duration of this period was ~9.1 d. We assume that this period coincides with the period of infectiousness and that this period encompasses both the incubation period and the immune response phase. According to our model, the mean duration of the incubation period is $5/k_L$ for both presymptomatic and asymptomatic individuals. Based on our earlier estimate that $k_L = 0.94$ d$^{-1}$, the mean duration of the incubation period is estimated as 5.3 d. Accordingly, the mean duration of the immune clearance phase for asymptomatic individuals is estimated as 9.1 d − 5.3 d = 3.8 d, and it follows that $c_A = \frac{1}{3.8 \text{ d}} \approx 0.26$ d$^{-1}$.

If $f_H \ll 1$, the parameter $c_I$ characterizes the duration of infectiousness of individuals who develop mild COVID-19 symptoms (i.e., symptoms not severe enough to require hospitalization). Wölfel et al. *(20)* attempted to isolate live virus from clinical throat swab and sputum samples collected from 9 patients at multiple time points after the onset of mild COVID-19 symptoms. Roughly 67%, 38%, and 0% of attempts to isolate virus were successful at 6, 8, and 10 d after infection, respectively. Assuming that a negative culture coincides with loss of infectiousness, we estimate that $c_I = -\frac{\ln(0.38)}{8 \text{ d}} \approx 0.12$ d$^{-1}$.

The parameters $f_R$ and $c_H$ characterize the hospital stays of the severely ill. These parameters affect predictions of COVID-19-caused deaths and hospital resource utilization but do not affect the predicted transmission dynamics, because we assume that hospitalized patients are effectively quarantined and do not contribute significantly to disease transmission, i.e., there is no $I_H$ term in $\phi_M$ or $\phi_P$ (see Equations (22) and (23)). The parameter $f_R$ is the fraction of hospitalized patients who recover, and the parameter $c_H$ characterizes the rate at which patients are discharged (as either recovered or dead). Richardson et al. *(19)* reported that the overall median length of hospital stay for 2,634 discharged patients (alive or dead) was 4.1 d. Thus, we estimate that $c_H = \frac{\ln(2)}{4.1 \text{ d}} \approx 0.17$ d$^{-1}$. Among the discharged patients, 553 (21%) died. Thus, we estimate that $f_R = 0.79$.

*Likelihood Function Used in Inference of Model Parameter Values*

We assume that the likelihood of a set of adjustable parameter values $\theta_F$ given a report of $\delta C_i$ new cases on calendar date $\mathcal{D}_i$, which we will denote as $\mathcal{L}_i(\theta_F; \delta C_i)$, is given by the following equation:

$$\mathcal{L}_i(\theta_F; \delta C_i) = \text{nbinom}(\delta C_i; r, p_i) = \binom{\delta C_i + r - 1}{\delta C_i - 1} p_i^r (1 - p_i)^{\delta C_i} \tag{29}$$

where $\delta C_i$ is a non-negative integer (the number of new cases reported), $i$ is an integer indicating the date $\mathcal{D}_i$ or the period $(t_i, t_{i+1})$; $\text{nbinom}(\delta C_i; r, p_i)$ is the probability mass function of the negative binomial distribution $\text{NB}(r, p_i)$, which has two parameters, $p_i \in [0,1]$ and $r > 0$; and $\theta_F$ is a model-dependent ordered set of feasible (i.e., allowable) values for the adjustable model parameters (e.g., $N$, $t_0$, $k$, and $\theta$ in the case of the curve-fitting model) augmented with a feasible value for $r$. Recall that $t_i \equiv t_0 + i$ d, where $t_0 > 0$ is 0000 hours of $\mathcal{D}_0$, the start date of the local epidemic. We take the dispersion parameter $r$ of $\text{NB}(r, p_i)$ to be date/time-independent and infer the value of $r$ jointly with the values of the model parameters. Because the deterministic models are designed to capture the mean behavior of the random reporting numbers, we have to impose the constraint $I(t_i, t_{i+1}) = \mathbb{E}[\text{NB}(r, p_i)] = r(1 - p_i)/p_i$, which leads to the date/time-dependent parameter $p_i$ of the negative binomial distribution:

$$p_i = \frac{r}{r + I(t_i, t_{i+1})}. \tag{30}$$

If $m + 1$ daily case reports are available, from date $\mathcal{D}_0$ to date $\mathcal{D}_m$, we assume that each likelihood $\mathcal{L}_i(\theta_F; \delta C_i)$ given by Equations (29) and (30) is independent. Thus, we have

$$\mathcal{L}(\theta_F; \{\delta C_i\}_{i=0}^m) = \prod_{i=0}^m \log \mathcal{L}_i(\theta_F; \delta C_i) \tag{31}$$

where $\mathcal{L}(\theta_F; \{\delta C_i\}_{i=0}^m)$ is the likelihood of $\theta_F$ given all available case reports $\{\delta C_i\}_{i=0}^m$. Recall that $\delta C_i$ is the number of new cases reported on date $\mathcal{D}_i$ and $I(t_i, t_{i+1})$ is the model prediction of $\delta C_i$. Furthermore, recall that $\theta_F$ is defined as a model-dependent ordered set of feasible adjustable model parameter values augmented with a feasible value for the likelihood function parameter $r$. The identity of $\theta_F$ depends on whether we are using Equation (31) to make inferences conditioned on the curve-fitting model or the compartmental model (i.e., we use Equation (31) in both cases but the identity of $\theta_F$ depends on the model being considered). The ordering of parameter values within the set $\theta_F$ is consistent but arbitrary.

*Bayesian Inference and Online Learning*

We chose the Bayesian inference framework to parametrize the models with uncertainty quantification. In Bayesian inference, given a set of data $D$, the probability of each set of the parameters, denoted in $\theta_F$ is constrained by the Bayes formula

$$\mathbb{P}\{\theta_F | D\} = \frac{\mathbb{P}\{D | \theta_F\} \, \mathbb{P}\{\theta_F\}}{\int_\Omega \mathbb{P}\{\theta_F' | D\} \, \mathbb{P}\{\theta_F'\} \, d\theta_F'}. \tag{32}$$

Here, the $\mathbb{P}\{\theta_F\}$ is the prior parameter distribution, which represents our belief of how the model parameters should distribute in the parameter space $\Omega$, and $\mathbb{P}\{D|\theta_F\}$ is the likelihood of the parameter set $\theta_F$ given the dataset $D$, that is, $\mathcal{L}(\theta_F; \{\delta C_i\}_{i=0}^m)$ in Eq. (31). In general, evaluating the posterior parameter distribution $\mathbb{P}\{\theta_F|D\}$ is a difficult computation, mainly because of the high-dimensional integration of the term $\int_\Omega \mathbb{P}\{\theta_F'|D\} \mathbb{P}\{\theta_F'\} \, d\theta_F'$, a term often referred to as the *evidence*. Thus, for high-dimensional models, one relies on Markov chain Monte Carlo (MCMC) techniques to sample the posterior parameter distribution $\mathbb{P}\{\theta_F|D\}$.

In contrast of many modeling analyses which focus on identifying the parameter distributions, we are interested in projections of the models, whose parameters are inferred by past data, into the future. To this end, we evaluate the model with a probabilistic parameter set distributed by the obtained posterior distribution $\mathbb{P}\{\theta_F|D\}$. Formally, we denote the prediction of the confirmed cases between future day $t_i$ and $t_{i+1}$ by our deterministic model with a set of parameters $\theta_F$ by $I(t_i, t_{i+1}; \theta_F)$. Recall that this deterministic prediction represents the mean of the fundamentally random new confirmed cases reported in a future interval $(t_i, t_{i+1})$. If there was only parametric uncertainty which propagates through the deterministic model, the confirmed cases reported in a future interval $(t_i, t_{i+1})$ would be distributed according to $\int_\Omega I(t_i, t_{i+1}; \theta_F) \, \mathbb{P}\{\theta_F|D\} \, d\theta_F$. However, there is also observation noise, which we model by a negative binomial distribution. The observation noise also needs to be injected into the prediction to quantify the full uncertainty. The full prediction accounting for parametric uncertainty is a random variable distributed according to

$$\int_\Omega \text{nbiom}\left(i; r, \frac{r}{r + I(t_i, t_{i+1}; \theta_F)}\right) \mathbb{P}\{\theta_F|D\} \, d\theta_F. \tag{33}$$

In practice, the above random variable is resampled from the posterior chain derived from the MCMC sampling. We denote the MCMC posterior chain by $\{\theta_F^{(1)}, \theta_F^{(2)} \ldots \theta_F^{(N)}\}$. We sample the posterior chain and denote the resampled parameter set by $\theta_F^s$ and the deterministic prediction of that resampled parameter in interval $(t_i, t_{i+1})$ by $I(t_i, t_{i+1}; \theta_F^s)$. Then, we generate a negative binomial random number with the first parameter of the negative binomial distribution set as $r_s/(r_s + I(t_{i+1}, t_i; \theta_F^s))$ where $r_s$ is the resampled $r$ which is also a free parameter in $\theta_F^s$ and is inferred in the MCMC. We repeat the resampling procedures and use the generated samples to compute the percentile of the past history and future prediction.

Our aim is to perform the Bayesian inference daily as soon as a new regional confirmed case number is reported. Although the Bayesian framework allows a sequential analysis, that the previous derived posterior distribution is used as a prior and the new inference problem involves only one new data point, in practice, such an analysis is difficult if the posterior distribution cannot be emulated or interpolated from the discrete posterior chain. Our analysis shows that in some regions, the posterior is far from Gaussian, making accurate interpolation or emulation difficult. Thus, we are not adopting this workflow, and instead, we perform the inference with all the data points collected up to the time of inference. Nevertheless, we warm-start the MCMC chain from the maximum a posteriori estimator estimated from the previous derived chain, and we also use the previously derived chain for estimating the optimal covariance matrix for the proposal of the normal symmetric random-walk Metropolis sampler. This approach allows an online learning of the optimal proposal which significantly reduces the mixing time.

*Technical Details of Approach and Numerical Methods Used in Bayesian Inference*

Because the variability of the data due to the regional and temporal differences, it is difficult to identify a universal sampling strategy (the proposal kernel). Thus, we adopted an adaptive Metropolis algorithm, specifically Algorithm 4 in Andrieu and Thoms *(21)* to accommodate the regional and temporal differences.

For all the model parameters, we assume their priors are uniformly distributed, whether they are proper or improper. Denote the 0:00 of the calendar date of the first confirmed case in a specific region by $t_{\text{first}}$ and the total population of that region by $S_0$. For the curve-fitting model, we assume the parameters are bounded by $N \in (0, S_0)$, $t_0 \in (t_{\text{first}} - 21, t_{\text{first}})$, $\mu \in (0, \infty)$, $k \in (0, \infty)$, $\theta \in (0, \infty)$. For the compartmental model, we assume that the parameters are bounded by $t_0 \in (0, \infty)$, $\sigma \in (t_0, \infty)$, $\beta \in (0, \infty)$, $\lambda \in (0,10)$, $f_P \in (0,1)$, $f_D \in (0,1)$. We assume $r \in (0, \infty)$ for both models. We adopted rejection-based sampling to assure the parameter values are sampled in the hypercube.

We start inference with an isotropic proposal kernel, that randomly perturbs the parameter values by independent Gaussian proposals whose standard deviations are set to be 5% of the parameter values. We carry out the standard Metropolis–Hastings algorithm for $5 \times 10^4$ iterations first to identify local minimum. Then, we turned on the adaptive Metropolis algorithm to calculate the covariance matrix on-the-fly, for another $5 \times 10^4$ iteration, when we turned on an on-the-fly learning for optimal proposed increment, i.e., $\lambda$ in Algorithm 4 of Andrieu and Thoms *(21)*. Because the weight for learning the proposed increment decays 1/iteration *(21)*, the proposed increment stabilizes after about $10^3$ more iterations. We began to collect the statistics from $1.5 \times 10^5$ iteration, until the simulation finishes at $6 \times 10^5$ iterations.

Except for the first time of the procedure (i.e., online learning and day-to-day operation), we warm start the simulation from the previously obtained best-fit (maximum a posteriori (MAP) estimator) and with the previously obtained covariance matrix and proposed increment. We carry out standard Metropolis–Hastings algorithms for $2.5 \times 10^4$ iterations first to identify a local minimum, noting that it is often not far away from the previously identified MAP. We then turn on the adaptive MCMC algorithm to calculate the covariance matrix on-the-fly, again for another $5 \times 10^4$ iteration. We then use another $2.5 \times 10^4$ iterations to calculate the optimal proposed increment. We start to collect statistic from $10^5$ iteration to $4 \times 10^5$ iteration when the simulation finishes.

**Appendix References**